\documentclass[acmsmall,nonacm]{acmart}
\usepackage{booktabs}   % professional tables (\toprule, \midrule, \bottomrule)
\usepackage{multirow}   % if needed for complex tables
\usepackage{array}      % better column control
\usepackage{tabularx}   % auto-adjusting table width (very useful in CSUR)
\usepackage{xcolor}     % for colored text (only if allowed by ACM)
\usepackage{pifont}     % for symbols (circles, ticks, crosses)
\usepackage{graphicx}   % scaling tables
\usepackage{adjustbox}  % best for ACM tables\
\usepackage{threeparttable} % add in preamble
\usepackage{wrapfig}

\usepackage{colortbl}
\newcommand{\gap}[1]{\textbf{(G#1)}}                       % research-gap IDs used by Section 6
\newcommand{\hm}[1]{\ifcase#1\relax\cellcolor{white}--\or\cellcolor{red!10}1\or\cellcolor{red!25}2\or\cellcolor{red!45}3\or\cellcolor{red!70}\textcolor{white}{4}\fi}

\newcommand{\myparagraph}[1]{\smallskip \noindent{\bf {#1}.}}
\usepackage{listings}
\usepackage{tcolorbox}

\usepackage[edges]{forest}
\definecolor{txInput}{HTML}{0072B2}   % blue
\definecolor{txModel}{HTML}{6A3D9A}   % purple
\definecolor{txCode}{HTML}{D55E00}    % vermillion
\definecolor{txAgent}{HTML}{00876C}   % bluish green
\definecolor{txProto}{HTML}{AA3377}   % magenta
\definecolor{txSupply}{HTML}{8C510A}  % brown
\definecolor{txInfra}{HTML}{B2182B}   % crimson
\definecolor{txHuman}{HTML}{4D4D4D}   % charcoal
\newcommand{\txsec}[1]{{\scriptsize\color{black!55}\S\ref{#1}}}
\forestset{
  cls/.style={draw=#1, text=#1, line width=1.1pt,
              for children={draw=#1, text=black, line width=0.7pt,
                            edge={draw=#1, line width=0.9pt}}},
}

\definecolor{cInk}{HTML}{1F2430}
\definecolor{cGry}{HTML}{9AA3AF}
\definecolor{cGryBg}{HTML}{F4F5F7}
\definecolor{cRed}{HTML}{B3261E}
\definecolor{cRedBg}{HTML}{FBEAE9}
\definecolor{cGrn}{HTML}{146C2E}
\definecolor{cGrnBg}{HTML}{E8F5EA}
\definecolor{cMut}{HTML}{6B7280}
\definecolor{txCross}{HTML}{1A1A1A}   % near-black, dashed = cross-cutting

\definecolor{txIndigo}{HTML}{332288}
\definecolor{txCyan}{HTML}{3A86A8}
\definecolor{txTeal}{HTML}{2F8F80}
\definecolor{txGreen}{HTML}{117733}
\definecolor{txOlive}{HTML}{8A8A2E}
\definecolor{txSand}{HTML}{A6842A}
\definecolor{txRose}{HTML}{B5525F}
\definecolor{txWine}{HTML}{882255}
\definecolor{txPurple}{HTML}{94408A}
\definecolor{txSlate}{HTML}{3E5C76}   % compound threats
\definecolor{txSpine}{HTML}{2E3F55}   % root and its connectors
\definecolor{txRef}{HTML}{6E6E6E}

\definecolor{txRef}{cmyk}{1,0.58,0,0.21}
 
\providecommand{\txsec}{}
\renewcommand{\txsec}[1]{~{\color{txRef}\S\ref{#1}}}

\acmJournal{CSUR}
\acmVolume{X}
\acmNumber{X}
\acmArticle{XXX}
\acmYear{2026}
\acmMonth{11}
\acmDOI{XXXXXXX.XXXXXXX}

\title{Threat Landscape of Vibe Coding and the Road to Secure Autonomy}
\author{Devki Nandan Jha}
\affiliation{
  \institution{School of Computing, Newcastle University}
  \country{UK}
}
\email{dev.jha@ncl.ac.uk}

\author{Shubh Anand}
\affiliation{
  \institution{School of Computing, Newcastle University}
  \country{UK}
}
\email{s.anand3@ncl.ac.uk}

\author{Karthick Keeranagere Krishnaiah}
\affiliation{
  \institution{School of Computing, Newcastle University}
  \country{UK}
}
\email{k.k.keeranagere-krishnaiah2@ncl.ac.uk}

\author{Roman Ahearn}
\affiliation{
  \institution{School of Computing, Newcastle University}
  \country{UK}
}
\email{r.l.ahearn2@ncl.ac.uk}

\author{Suhas Nale}
\affiliation{
  \institution{School of Computing, Newcastle University}
  \country{UK}
}
\email{s.s.nale2@ncl.ac.uk}

\begin{abstract}
Vibe coding is reshaping software development, enabling faster implementation and allowing individuals with limited programming experience to describe requirements in natural language while AI agents write, test, and deploy the code. As these agents evolve into autonomous systems that execute commands, manage dependencies, access cloud resources, and interact with other agents, their security risks extend beyond incorrect code to actions with direct consequences for software systems. Although research on these risks is expanding, it remains fragmented across areas including prompt injection, insecure code generation, supply-chain attacks, and human factors, with limited consensus on terminology, threat assumptions, and evaluation. This survey consolidates emerging research through the lens of the vibe coding workflow. We establish a reference architecture, characterise six levels of agent autonomy, and develop a threat model centred on five trust boundaries spanning developers, AI models, tools, cloud infrastructure, and the wider software ecosystem. Building on this foundation, we propose a taxonomy of twenty-one threats across nine classes. For each threat, we review the available evidence, assess proposed defences, and identify limitations. Our analysis highlights that risks increase with agents' operational authority, while most defences rely on model judgement and address isolated threats. Finally, we identify eleven research challenges, including enforceable limits on agent authority and shared benchmarks and methodologies for assessing vibe coding security.

\end{abstract}



\usepackage{tikz}
\usepackage{fontawesome5} % Added for scalable, professional vector icons
\usetikzlibrary{shapes.geometric, arrows.meta, positioning, fit, calc, backgrounds}

\usepackage[nointegrals]{wasysym}
\newcommand{\hbfull}{\CIRCLE}
\newcommand{\hbhalf}{\LEFTcircle}
\newcommand{\hbnone}{\Circle}

\begin{document}
\maketitle

\section{Introduction} \label{sec:intro}
Software development is undergoing a fundamental transformation driven by the rapid advancement of large language models (LLMs) \cite{LLMFSE, ARR}. What began as code completion has rapidly evolved into a new programming paradigm in which natural language becomes the main interface. Developers describe what they want in plain language, and an AI system translates the intent into executable code. This paradigm, widely referred to as vibe coding, shifts the developer's role from writing code to stating goals, reviewing generated artefacts, and guiding iterative refinement \cite{fawzy2026vibe, VCA}. It also opens software development to people with little or no programming experience, who build and publish complete applications through conversation only.

The evolution of vibe coding reflects the growing capabilities and autonomy of LLM-based development systems. Unlike earlier tools designed primarily as autocomplete assistants, recent systems, such as GitHub Copilot\footnote{https://github.com/features/copilot}, Claude Code\footnote{https://claude.com/product/claude-code}, Devin\footnote{https://devin.ai/}, and Codex\footnote{https://openai.com/codex/}, can inspect repositories, plan multi-step tasks, invoke development tools, install dependencies, execute code, and open pull requests with limited human guidance. These capabilities often require read and write access to source repositories, permission to execute shell commands and install packages, integration with CI/CD pipelines, and, in some deployments, access to cloud infrastructure and production services \cite{jiang2026survey, zhang2026agentauditsecurityanalysis}. This shift changes the security assumptions of software development. While early completion tools primarily produced suggestions, autonomous agents can modify dependencies, execute code in build environments, and introduce changes that may bypass human review. They are therefore no longer merely productivity tools but security-relevant components of the software supply chain \cite{ladisa2023sok}, where errors and vulnerabilities can translate directly into harmful actions.

Existing studies highlights the growing evidence of these risks. Developers using AI assistants have been found to produce less secure code while feeling more confident in its security \cite{perry2023users}, and models can recommend software packages that do not exist \cite{sparacklen2025package}. An audit of 200 vibe-coded applications found at least one vulnerability in more than 90\% of them \cite{deng2026understanding}. Despite growing research interest, the literature remains fragmented across insecure code generation \cite{claudia2024systematic, kharma2026empirical}, prompt injection \cite{10.1145/3605764.3623985, debenedetti2024agentdojo}, package hallucination \cite{sparacklen2025package}, unsafe execution environments \cite{rashidi2026balkanization}, and human overtrust in generated code \cite{GLAS2026104842, sabouri2025trust}. These studies use different terminologies, threat assumptions and evaluation methods, making it difficult to understand how vulnerabilities across the development workflow interact and contribute to broader security risks.

\myparagraph{Related Surveys and Paper Positioning}
The survey literature shows similar fragmentation. One group of surveys studies LLM-based code generation and the vibe coding paradigm, covering model capabilities, development practice and human and AI collaboration, and treats security as one open challenge among many~\cite{fawzy2026vibe, VCA, krasniqi2025se}. A second group examines the security of AI-generated code, either by reviewing its weaknesses or by surveying the use of LLMs to find and repair vulnerabilities~\cite{claudia2024systematic, ramirez2025state, kaniewski2026systematic}. A third group surveys threats to LLMs and LLM agents in general~\cite{das2025security, wang2026unique, wu2026comparative}. This group includes a layered model of the agent attack surface~\cite{chu2026systematic} and a systematisation of execution security for coding agents~\cite{rashidi2026balkanization}. Other surveys go deep into single areas such as prompt injection~\cite{FERRAG2026353, Geng2026}, the software supply chain~\cite{williams2025research} and developer trust in AI tools~\cite{baltes2026need, ehsani2026faster}. These surveys offer valuable insights into specific aspects of AI-assisted development, but they do not capture how security risks emerge and interact across the entire vibe coding workflow, from the initial prompt to production. To the best of our knowledge, no existing survey systematically maps threats to the level of autonomy granted to the agent or examines how vulnerabilities across different stages of the workflow can combine to form a single, end-to-end attack. Table~\ref{tab:surveys} compares this survey with the closest existing work.

\begin{table}[t] 
\caption{Comparison of recent surveys. \textbf{VC} = specific to vibe coding; \textbf{TM} = unified threat or trust model; \textbf{TX} = multi-category vulnerability taxonomy; \textbf{AU} = analysis linked to levels of autonomy; \textbf{MT} = systematic coverage of mitigations. Coverage: \protect\hbnone{} not addressed, \protect\hbhalf{} partially addressed, \protect\hbfull{} fully addressed.}
\label{tab:surveys}
\scriptsize
\setlength{\tabcolsep}{3pt}
\renewcommand{\arraystretch}{1.15}
% \begin{tabular}{@{}p{2.8cm}p{3.6cm}ccccc@{}}
\begin{tabularx}{\columnwidth}{@{}p{3.8cm}>{\raggedright\arraybackslash}Xccccc@{}}
\toprule
\textbf{Paper} & \textbf{Primary focus} &
\textbf{VC} & \textbf{TM} & \textbf{TX} & \textbf{AU} & \textbf{MT} \\
\midrule
Fawzy et al. ~\cite{fawzy2026vibe}       & Vibe Coding motivation and challenges      & \hbfull & \hbnone & \hbnone & \hbhalf & \hbnone \\
Meske et al. ~\cite{VCA}       & Vibe Coding as intent mediation in software development         & \hbfull & \hbhalf & \hbnone & \hbnone & \hbnone \\
Krasniqi et al. ~\cite{krasniqi2025se}       & LLM risks for software engineering    & \hbnone & \hbhalf & \hbhalf & \hbnone & \hbhalf \\
Kaniewski et al. ~\cite{kaniewski2026systematic}       & Detecting software vulnerabilities with LLM    & \hbnone & \hbhalf & \hbhalf & \hbnone & \hbfull \\
Negri-Ribalta et al. ~\cite{claudia2024systematic}       & Security of AI generated code    & \hbnone & \hbnone & \hbhalf & \hbnone & \hbhalf \\
Criollo Ram{\'i}rez~\cite{ramirez2025state}               & Security of AI generated code            & \hbnone & \hbnone & \hbhalf & \hbnone & \hbhalf \\
Das et al. ~\cite{das2025security}        & LLM security and privacy challenges           & \hbnone & \hbhalf & \hbhalf & \hbnone & \hbfull \\
Chu et al. ~\cite{chu2026systematic}        & Security threats \& defenses in LLM-based AI agents           & \hbnone & \hbhalf & \hbhalf & \hbhalf & \hbfull \\
Williams et al. ~\cite{williams2025research}       & Software supply chain security      & \hbnone & \hbfull & \hbfull & \hbnone & \hbfull \\
Ferrag et al. ~\cite{FERRAG2026353}, Geng et al. \cite{Geng2026}      & Prompt injection attacks in LLM system            & \hbnone & \hbhalf & \hbfull & \hbnone & \hbfull \\
Ehsani et al. ~\cite{ehsani2026faster}, Baltes et al. \cite{baltes2026need}      & Human factor and trust in LLM-based software engineering        & \hbnone & \hbfull & \hbhalf & \hbhalf & \hbfull \\
Wang et al. ~\cite{wang2026unique}        & LLM security and privacy threats            & \hbnone & \hbhalf & \hbhalf & \hbnone & \hbfull \\
Wu et al. ~\cite{wu2026comparative}       & Security risks in LLMs \& agents          & \hbnone & \hbhalf & \hbfull & \hbhalf & \hbfull \\
Rashidi et al. ~\cite{rashidi2026balkanization}       & Coding agents execution security intelligence        & \hbhalf & \hbhalf & \hbhalf & \hbnone & \hbfull \\
\midrule
\textbf{This survey}                    & \textbf{Vibe coding security}       & \hbfull & \hbfull & \hbfull & \hbfull & \hbfull \\
\bottomrule
% \end{tabular}
\end{tabularx}
\end{table}

\myparagraph{Contributions} 
This survey is organised around the following three research questions:
\begin{itemize}
    \item[\textit{RQ1}:] How can the vulnerabilities of vibe coding systems be organised into a unified taxonomy that links each threat to the components it affects and the trust boundaries it crosses?
    \item[\textit{RQ2}:] What defence mechanisms, benchmarks, and evaluation methodologies have been proposed, and what limitations remain in existing approaches?
    \item[\textit{RQ3}:] What are the most critical open challenges that must be addressed to make vibe coding secure at higher levels of autonomy?
\end{itemize}

To answer these questions, this paper makes four main contributions. \emph{(1)} We describe a reference architecture for vibe coding systems and identify six levels of autonomy (Section~\ref{sec:background}), which inform a threat model comprising five trust boundaries across which data and authority cross (Section~\ref{sec:threat-model}). \emph{(2)} We propose a threat taxonomy of nine threat classes, organised according to the system architecture (Section~\ref{sec:taxonomy}). \emph{(3)} For each threat, we review the available evidence, proposed defences and their evaluation, and identify twenty-one research gaps (Section~\ref{sec:taxonomy}). \emph{(4)} We consolidate these gaps into eleven overarching research challenges and outline concrete directions for future work (Section~\ref{sec:challenges}).

\myparagraph{Survey Methodology}
We followed a structured literature review to identify research on the security of vibe coding and AI-assisted software development. We searched the ACM Digital Library, IEEE Xplore, USENIX, Scopus and Google Scholar, and also included arXiv because much of this field first appears as preprints. We focused on work published from 2021, when AI code completion entered mainstream use, to August 2026. The search combined terms for vibe coding and AI coding tools, terms for the agentic components these tools rely on, and terms for security (Appendix~\ref{sec:AppendixA}), returning 1211 articles. After removing duplicates and screening titles and abstracts, we read the abstracts of 546 papers and kept those that report an attack, a defence, an empirical measurement or a benchmark concerning the security of AI-assisted or agentic software development. We excluded work on general LLM safety with no software development setting and position papers without supporting evidence, leading to 123 papers for full reading. Backwards and forward snowballing from the selected papers added further studies, giving 142 primary studies in total.  

\myparagraph{Outline} The remainder of this paper is organised as follows. Section \ref{sec:background} introduces the background of vibe coding systems. Section \ref{sec:threat-model} presents a unified threat model. Section \ref{sec:taxonomy} develops the proposed vulnerability taxonomy, discussing each vulnerability with the existing defence mechanisms and challenges. Section \ref{sec:challenges} discusses future research directions. Finally, Section \ref{sec:conclusion} concludes the paper.

\section{Vibe Coding} \label{sec:background}
Vibe coding is a paradigm of software development in which developers express their software requirements primarily through natural-language prompts and delegate much of the implementation to LLM-based systems. The term was introduced by Andrej Karpathy in February 2025 to describe a style of programming in which developers accept AI-generated code with limited inspection or understanding of its underlying implementation\footnote{https://x.com/karpathy/status/1886192184808149383}. In this original sense, vibe coding differs from conventional AI-assisted programming, where developers remain actively involved in reviewing, testing, and understanding the generated code \cite{willison2025}. However, the term has since broadened to include developing software through natural-language interactions with AI, a shift reflected in Collins Dictionary naming vibe coding its Word of the Year for 2025\footnote{https://www.collinsdictionary.com/woty}. 

Recent research also examines vibe coding as a shift in how developers communicate their intentions and collaborate with AI systems~\cite{fawzy2026vibe, VCA, garfinkel2026acm}. In this survey, we adopt this broader view and use vibe coding to refer to software development in which natural language serves as the primary interface and AI systems undertake an increasing share of implementation, with the degree of human review and control varying across tools and workflows. This approach spans both professional developers, who use AI coding agents within editors and terminals, and non-programmers, who build applications through conversational platforms such as Bolt\footnote{https://bolt.new/} and Replit\footnote{https://replit.com/}. This distinction is particularly relevant to security, as experienced developers may be able to inspect and validate generated code, while users with limited programming knowledge may have difficulty identifying vulnerabilities in applications they create. This section provides the background necessary for understanding the remainder of the survey.

\subsection{Evolution and Autonomy of Vibe Coding}\label{sec:autonomy}
% \begin{wrapfigure}{r}{0.6\textwidth}
%   \begin{center}
%     \vspace{-2em}
%     \includegraphics[width=0.70\textwidth]{VibeCoding Evolution1.pdf}
%   \end{center}
%    \vspace{-1em}
%   \caption{Vibe Coding Evolution}
%   \vspace{-1em}
% \end{wrapfigure}

The emergence of vibe coding has been driven by the continuous evolution of AI-assisted programming tools. Early tools such as IntelliSense completed individual tokens based on the current file \cite{microsoft_intellisense}, while later IDEs used the wider project to improve suggestions. Machine learning changed this around 2018, when TabNine \cite{tabnine2020} and then GitHub Copilot$^1$ began to suggest whole functions learned from public code. This development marked the transition from predictive code completion to AI-assisted code generation. The release of ChatGPT\footnote{https://chatgpt.com/} in late 2022 made it possible to ask for code in plain language. Repository-aware development environments such as Cursor\footnote{https://cursor.com/} further extended this capability by providing the model with access to project-wide context, enabling coordinated modifications across multiple files and software components. The subsequent arrival of agentic workflows, instantiated by systems such as Devin$^3$, OpenHands \cite{wang2024openhands}, SWE-agent \cite{yang2024sweagent} and Claude Code$^2$, moved AI assistance beyond generation into action. The Model Context Protocol\footnote{https://modelcontextprotocol.io/specification/2024-11-05}, released in November 2024, provided a common way for these agents to access external tools. More recently, coding agents have begun opening pull requests in the background, while app builders can create and deploy complete applications. Across these changes, assistants have gained greater access to the development process, while developers have become less involved in checking each individual action.

This shift in responsibility can be described through six levels of autonomy, summarised in Table~\ref{tab:levels}. Drawing on the user-centred levels~\cite{feng2025levels} and the driving automation levels \cite{sae2021}, we define each level by what the agent can do without permission and what the human must still approve. Our levels focus on software development and reflect the agent's authority over the code repository, available tools and deployment environment. Two levels mark important security changes: \textit{(i)} at L3 the agent starts to execute actions rather than only propose them, meaning that a flaw can affect the developer's machine or a build server and \textit{(ii)} at L4, the human no longer approves each step but reviews the final result, such as a pull request, allowing flaws to reach shared systems before they are checked. These levels describe how an agent is used rather than the product itself. The same coding agent may operate at L2 when each change requires approval or at L4 when it works unattended in the cloud.

\begin{table}[t]
\centering
\caption{Levels of autonomy in vibe coding systems. Each level is defined by the authority the agent holds and by what the human still approves.}
\label{tab:levels}
\scriptsize
\setlength{\tabcolsep}{4pt}
\renewcommand{\arraystretch}{1.15}
\begin{tabularx}{\textwidth}{@{}l l >{\raggedright\arraybackslash}X >{\raggedright\arraybackslash}X >{\raggedright\arraybackslash}X@{}}
\toprule
\textbf{Level} & \textbf{Name} & \textbf{Agent Operation} & \textbf{Human Approval} & \textbf{Reach} \\
\midrule
L0 & Autocomplete & Suggest the next tokens or lines in the open file & Every suggestion, before it is inserted & The line being written \\
L1 & Guided generation & Produce code fragments in answer to a prompt & Whether to copy each fragment into the project & The fragment the developer adopts \\
L2 & Multi-file editing & Read the repository and propose edits across files & Each set of changes, shown as a diff & The repository, once a diff is accepted \\
L3 & Tool-using agent & Run commands, tests, package installs and external tools & Individual commands or risky steps, while supervising the session & The developer's machine, build server and connected services \\
L4 & Autonomous agent & Plan, implement, test and submit changes without step-by-step approval & The finished outcome, such as a pull request or deployed app & Shared infrastructure and production \\
L5 & Self-directed agent & Start maintenance and improvement tasks on its own & The policies that govern the agent, and exceptions & Anything the agent's policy allows \\
\bottomrule
\end{tabularx}
\vspace{-2em}
\end{table}

Throughout the paper, we use the term \emph{vibe coding system} for the whole set of software that turns a developer's request into working code, which includes the host application, the model behind it, the tools it can call and the platform that serves it. When such a system can act on its environment, we call it an \emph{agent}.

\subsection{Vibe Coding: System Architecture} \label{sec:architecture}

\begin{figure}[t!]
    \centering
    \includegraphics[width=0.85\linewidth]{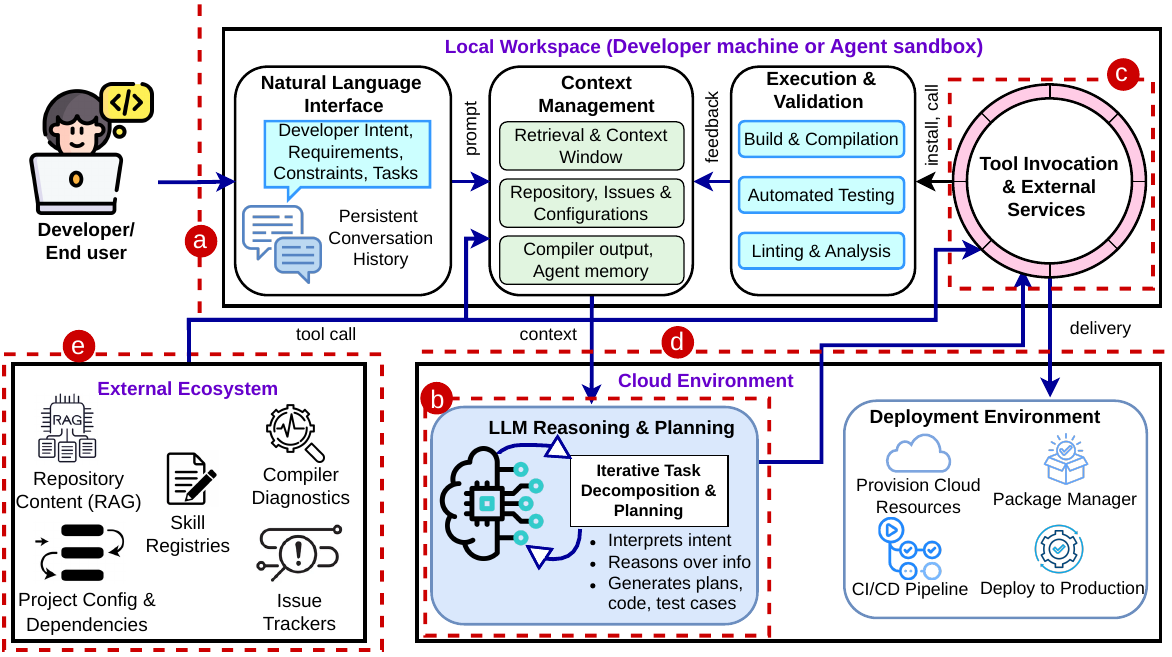}
    \vspace{-0.5em}
    \caption{Vibe Coding Architecture. Red dashed lines mark the five trust boundaries: (a) human-agent, (b) agent-model, (c) agent-tool, (d) local-cloud, and (e) agent-ecosystem.}
    \label{fig:vc-achitecture}
    \vspace{-1em}
\end{figure}

Although individual platforms differ in implementation, typical vibe coding systems follow a common architectural pattern, as shown in Figure \ref{fig:vc-achitecture}. It consists of six components that work together across the developer's local workspace, a cloud provider and external services. 

\paragraph{I. Natural Language Interface:} Each session begins with a developer specifying a goal or a task and some constraints using plain language. Many tools also read \emph{instruction files}, such as \texttt{AGENTS.md} or editor rules, which guide the whole session. The system keeps the conversation, allowing historical prompts and responses to influence future reasoning.

\paragraph{II. Context Management:} Before calling the LLM, the host prepares its \emph{context window}, the block of text the model reads at one time. This includes the developer's request and relevant project material, such as source files, issues, configuration and compiler diagnostics. As modern repositories frequently exceed the context window, systems use retrieval-augmented generation (RAG) or similar retrieval mechanisms to identify and add only the most relevant artefacts to the prompt. Some systems also maintain \emph{agent memory}, which stores notes and past interactions between sessions. The information included in this context shapes the model's output and can affect every later action.

\paragraph{III. LLM Reasoning and Planning:} The model reads the assembled context, interprets the developer's intent, reasons over the available information, and generates implementation plans, source code, explanations, test cases, configuration changes, or repository modifications. In agentic systems, it also includes task decomposition, multi-step planning, memory management, and iterative decision-making before any action is executed. 

\paragraph{IV. Tool Invocation and External Services:} To act on the desired plan, the model makes \emph{tool calls}, which the host carries out on its behalf. This may include running a shell command, installing a package or querying a database. The Model Context Protocol (MCP)$^11$ standardises how tools are described and called, so third-party \emph{MCP servers} and skills can add new capabilities to any compatible agent. In multi-agent systems, agents can also delegate tasks to one another through protocols such as Agent2Agent (A2A)\footnote{https://agent2agent.info/specification/}.

\paragraph{V. Execution and Validation:} The system applies the generated changes and checks them through builds, tests, linting and other analysis. Error messages, failing tests and logs are fed back into the context, allowing the agent to revise its work and try again iteratively. This feedback loop helps modern agents handle tasks over many steps with limited human input, enabling autonomous software engineering workflows.

\paragraph{VI. Deployment Environment:} At higher-autonomy levels, the loop extends beyond software development into operational environments. Agents trigger pull requests, edit CI/CD pipelines and infrastructure-as-code (IaC) templates, provision cloud resources, and deploy software to production systems. 

One further part of the architecture sits outside the developer's direct control. \emph{External ecosystem} supplies the repository contents for retrieval, project configuration, dependencies, and skill registers. It may also contain the compiler diagnostics for further analysis.

\section{Vibe Coding Threat Model} \label{sec:threat-model}
While the architectural components of vibe coding as discussed in Section \ref{sec:architecture} enable powerful development capabilities, they also introduce new trust relationships that are largely absent from traditional software engineering. Developers increasingly rely on AI systems to interpret natural-language requirements, retrieve contextual information, reason over software engineering tasks, and modify software repositories. The security of vibe coding thus depends not only on protecting individual components, but also on preserving the trust assumptions that exist between them. To provide a common analytical foundation for the remainder of this survey, we propose a threat model based on three complementary dimensions: \emph{information}, \emph{authority}, and \emph{autonomy}. Together, these dimensions explain how attacks arise, how they propagate across the software engineering workflow, and why increasingly autonomous coding agents introduce fundamentally different security risks compared with traditional development tools.

% To establish a common foundation for the remainder of this survey, we define a unified threat model that addresses four fundamental questions: \emph{Q1.} What trust assumptions underpin modern vibe coding systems? \emph{Q2.} Who are the potential adversaries? \dev{Who can violate these trust assumptions?} \emph{Q3.} How can trust violations propagate through the software development lifecycle? \dev{\emph{Q4.} What security objectives should a secure vibe coding system satisfy?} The answers to these questions provide the conceptual basis for the vulnerability taxonomy presented in the following section.

\subsection{Security Dimensions of Vibe Coding}\label{sec:dimensions}
The security of vibe coding extends beyond the generated source code. We group security-relevant assets and capabilities into three complementary dimensions as given in the following. 

\emph{I. Information} represents everything that influences the reasoning process of the coding agent. This includes developer prompts, conversation history, retrieved repository context, documentation, dependency metadata, execution feedback, and persistent agent memory. The integrity, authenticity, and provenance of this information directly affect the quality and security of the generated software.

\emph{II. Operational Authority} describes the actions that an autonomous vibe coding system is permitted to perform. As the system possesses capabilities extending beyond source-code generation, improving developers' productivity, it also expands the operational consequences of incorrect or adversarial decisions. As the permissions increase, these systems increasingly become security-relevant actors rather than passive development assistants.

\emph{III. Autonomy} represents the degree to which software engineering decisions are delegated to the AI system. As discussed in Section \ref{sec:autonomy}, autonomy ranges from simple code completion to fully autonomous software engineering agents capable of planning, implementing, testing, debugging, and deploying software with limited human intervention. Increasing autonomy reduces the frequency of direct human validation while enabling the agent to make progressively more consequential decisions. Moreover, the security implications of a vulnerability depend not only on the underlying issues but also on the level of autonomy granted to the system.

\subsection{Adversary Model}
The threat landscape of vibe coding involves adversaries operating at multiple levels of the software engineering ecosystem. We define five broad adversary classes based on their capabilities rather than the specific components they target.

\emph{I. External Adversaries} are attackers with no legitimate access to the vibe coding system or the project it operates on. Given the context, these adversaries typically act through the inputs the system is designed to trust, e.g., natural language prompts, code comments, README files, issue descriptions, or entire repositories crafted to influence the agent's behaviour. As vibe coding agents are built to follow natural language instructions, this class of adversaries can influence system behaviour without ever gaining direct access to it. 
% Goal: steer code generation, leak information, or trigger unintended actions purely through crafted input.

\emph{II. Compromised Insider} can be a developer, reviewer, or team member with legitimate access to the vibe coding environment who misuses that access, either deliberately or after their account or credentials have been compromised. They are particularly significant in vibe coding workflows as the wrong access can approve unsafe or malicious changes with minimal scrutiny. 
% Goal: insert vulnerabilities, exfiltrate data, or bypass review processes using access that the system treats as trustworthy.

\emph{III. Compromised Agents} act against the developer's interest, either because they have been manipulated by an external adversary, for example through prompt injection, or because their own generation and decision-making are unreliable in a given context. This adversary class is specific to vibe coding, where the agent is trusted with tasks that range from writing code to executing it.
% Goal: varies with the situation, but commonly results in unsafe code, unauthorized tool use, or actions the developer did not intend.

\emph{IV. Third-party Adversary} controls or compromises an external API, tool, or service that the vibe coding agent has been granted access to while completing a task, such as a code execution sandbox, a deployment service, or a data source. 
% Goal: intercept data passed to that service, or exploit the agent's trust in the tool to perform actions the developer did not authorize.

\emph{V. Supply-chain Adversary} targets the broader ecosystem that a vibe coding system depends on, including the underlying language model, its training data, plugins or extensions, and third-party packages the agent may recommend or install. Since vibe coding systems often pull in dependencies automatically as part of generating a working solution, a single upstream compromise can reach many projects without any additional action from the attacker. Table \ref{tab:adversary} summarises adversary classes, along with their capabilities and goals in a vibe coding setting.
% Goal: introduce a compromise once, upstream, so that it propagates automatically to downstream users of the system.

\begin{table}[]
\scriptsize
% \centering
\caption{Adversary classes in vibe coding systems}
\vspace{-1em}
\label{tab:adversary}
% \resizebox{\textwidth}{!}{%
\begin{tabular}{|c|c|c|}
\hline
\textbf{Adversary Class} & \textbf{Capabilities}                                                                                       & \textbf{Goals}                                                                           \\ \hline
External Adversary       & \begin{tabular}[c]{@{}c@{}}Crafts malicious prompts, files, or repositories the \\agent reads\end{tabular} & \begin{tabular}[c]{@{}c@{}}Influence generation, trigger \\ unwanted action\end{tabular} \\ \hline
Compromised Insider      & Legitimate but misused access                                                                               & Insert vulnerabilities, exfiltrate data                                                  \\ \hline
Compromised Agent        & \begin{tabular}[c]{@{}c@{}}Acts autonomously with granted permissions\end{tabular}                      & Unsafe code, unauthorised tool use                                                       \\ \hline
Third-party Adversary    & \begin{tabular}[c]{@{}c@{}}Controls an external API or tool the agent uses\end{tabular}                   & Data interception, unauthorised action                                                   \\ \hline
Supply-chain Adversary   & \begin{tabular}[c]{@{}c@{}}Compromises the model, training data, or dependencies\end{tabular}         & Persistent, wide-scale compromise                                                        \\ \hline
\end{tabular}%
% }
\vspace{-1em}
\end{table}

\subsection{Trust Boundary and Threat Propagation}\label{sec:threat-propagation}
Threats in a vibe coding system are rarely confined to a single point. A vulnerability may originate within a single layer of the architecture (described in Section \ref{sec:architecture}), independent of any external interference, or it may enter at one point and propagate across several components before causing harm. This propagation typically occurs as information, authority, or control crosses the interfaces between different system components. These interfaces, trust boundaries, represent a point where one entity must rely on the integrity of data or actions received from another, making them critical locations for understanding how vulnerabilities can spread throughout the system. Unlike traditional software development, where developers perform most engineering tasks directly, vibe coding delegates substantial responsibility to AI agents that continuously exchange information with external systems. Consequently, every transfer of information or authority becomes a potential point of attack. We identify five such boundaries in vibe coding systems. \emph{(a) The human-agent boundary}, where a developer's intent, prompts, or approval passes to the agent. \emph{(b) The agent-model boundary}, where prompts and context are sent to the underlying LM, and its output is returned. \emph{(c) The agent-tool boundary}, where the agent invokes external tools, compilers, or shell commands. \emph{(d) The local-cloud boundary}, where code, data, or credentials move between a developer's machine and a cloud-hosted service. \emph{(e) The agent-ecosystem boundary}, where the agent pulls in external dependencies, packages, or plugins.

Each of these boundaries rests on an implicit trust assumption that the party on the other side of the interface is behaving as expected. Developers assume that prompts accurately capture their intent, retrieved context and external knowledge are authentic, repositories faithfully represent the current software state, and LM-generated outputs satisfy both functional and security requirements. Similarly, coding agents assume that external tools, package repositories, MCP services, execution environments, and cloud platforms behave as expected and return trustworthy results. These assumptions can be violated by any of the adversary classes described earlier, the effect is rarely limited to the boundary at which it occurred. As vibe coding agents chain multiple boundaries within a single task, authority or data that has been compromised at one interface is often carried forward and trusted again at the next, allowing a single violation to escalate into a broader failure. This perspective provides the foundation for understanding the diverse classes of vulnerabilities surveyed throughout this paper.

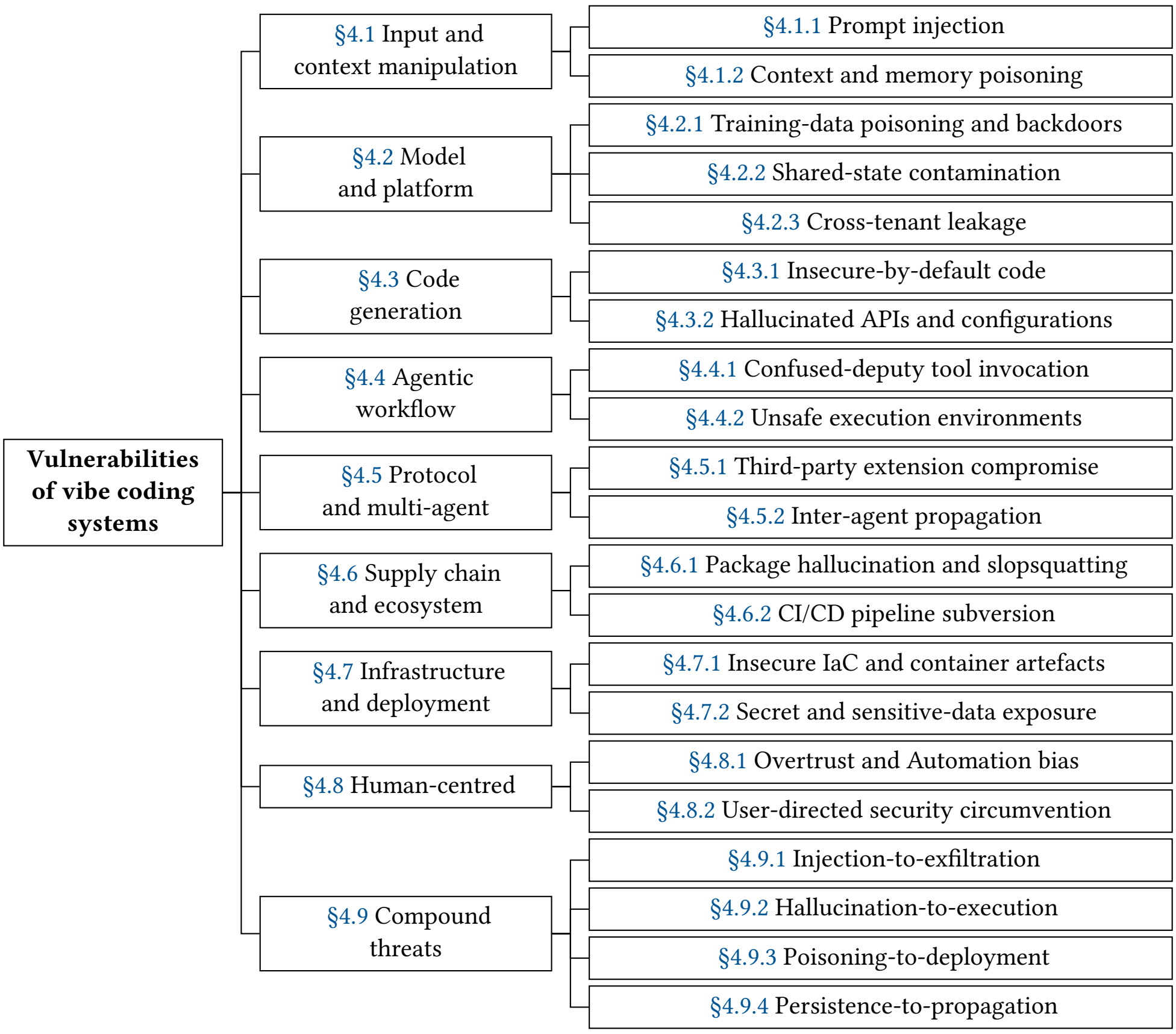
\begin{figure}[t]
\centering
\begin{forest}
  forked edges,
  for tree={
    grow'=east,
    draw=black, line width=0.4pt,       % thin black, square corners
    align=center,
    inner xsep=3pt, inner ysep=1.6pt,
    l sep=4mm, s sep=0.6mm, fork sep=2mm,
    edge={draw=black, line width=0.4pt},
    font=\footnotesize,
    anchor=west, child anchor=west, parent anchor=east,
  },
  where level=0{font=\footnotesize\bfseries, minimum width=2.4cm}{},
  where level=1{minimum width=3.2cm, s sep=0.6mm}{},
  where level=2{minimum width=6.4cm}{},
[{Vulnerabilities\\of vibe coding\\systems}
  [{\txsec{sec:tx-input} Input and \\ context manipulation}
    [{\txsec{sec:tx-pi} Prompt injection}]
    [{\txsec{sec:context-poisoning} Context and memory poisoning}]
  ]
  [{\txsec{sec:tx-model} Model\\and platform}
    [{\txsec{sec:tx-backdoor} Training-data poisoning and backdoors}]
    [{\txsec{sec:tx-shared} Shared-state contamination}]
    [{\txsec{sec:tx-leakage} Cross-tenant leakage}]
  ]
  [{\txsec{sec:tx-codegen} Code\\generation}
    [{\txsec{sec:tx-insecure-code} Insecure-by-default code}]
    [{\txsec{sec:tx-halluc} Hallucinated APIs and configurations}]
  ]
  [{\txsec{sec:tx-agentic} Agentic\\workflow}
    [{\txsec{sec:tx-deputy} Confused-deputy tool invocation}]
    [{\txsec{sec:tx-exec} Unsafe execution environments}]
  ]
  [{\txsec{sec:tx-protocol} Protocol\\and multi-agent}
    [{\txsec{sec:tx-mcp} Third-party extension compromise}]
    [{\txsec{sec:tx-a2a} Inter-agent propagation}]
  ]
  [{\txsec{sec:tx-supply} Supply chain\\and ecosystem}
    [{\txsec{sec:tx-slop} Package hallucination and slopsquatting}]
    [{\txsec{sec:tx-cicd} CI/CD pipeline subversion}]
  ]
  [{\txsec{sec:tx-infra} Infrastructure\\and deployment}
    [{\txsec{sec:iac} Insecure IaC and container artefacts}]
    [{\txsec{sec:tx-secrets} Secret and sensitive-data exposure}]
  ]
  [{\txsec{sec:tx-human} Human-centred}
    [{\txsec{sec:tx-overtrust} Overtrust and Automation bias}]
    [{\txsec{sec:tx-circumvent} User-directed security circumvention}]
  ]
  [{\txsec{sec:tx-compound} Compound\\threats}
    [{\txsec{sec:tx-chain-exfil} Injection-to-exfiltration}]
    [{\txsec{sec:tx-chain-exec} Hallucination-to-execution}]
    [{\txsec{sec:tx-chain-deploy} Poisoning-to-deployment}]
    [{\txsec{sec:tx-chain-persist} Persistence-to-propagation}]
  ]
]
\end{forest}
\caption{Taxonomy of vulnerabilities in vibe coding systems.}
\label{fig:taxonomy}
\vspace{-1em}
\end{figure}

\section{Vulnerability Taxonomy for Vibe Coding Systems}\label{sec:taxonomy}
This section presents a taxonomy of the vulnerabilities of vibe coding systems. Figure~\ref{fig:taxonomy} shows its nine classes and twenty-one threats.

\subsection{Input and Context Manipulation Threats}\label{sec:tx-input}
In conventional software engineering, system inputs are structured, typed and sanitised by deterministic routines. In vibe coding, the primary input is natural language, and the context window concatenates developer intent, retrieved artefacts and tool output into a single token sequence with no enforced separation between instructions and data. Two threats exploit this property: Prompt Injection corrupts the \emph{instructions} the agent follows, whereas Context and Memory poisoning corrupts the \emph{premises} from which it reasons. Figure~\ref{fig:threat4.1} contrasts the two types of threats.

\subsubsection{Prompt Injection}\label{sec:tx-pi}
Prompt injection occurs when adversary-controlled content entering the context window is interpreted as an instruction rather than as data. Injection can be \emph{direct} when the adversarial text is supplied through the prompt itself, or \emph{indirect} when it arrives through content the agent retrieves while performing a task. Greshake et al.~\cite{10.1145/3605764.3623985} showed that indirect injection can hijack LLM-integrated applications without any access to the system prompt, establishing it as an architectural rather than a prompt-engineering weakness. In vibe coding, the indirect channel is native to the workflow, as README files, issue descriptions, code comments, package metadata, web pages and tool outputs are all routinely ingested across the agent-model boundary, while the consequences are realised at the agent-tool boundary.

\begin{figure}[t!]
    \centering
    \includegraphics[width=0.82\linewidth]{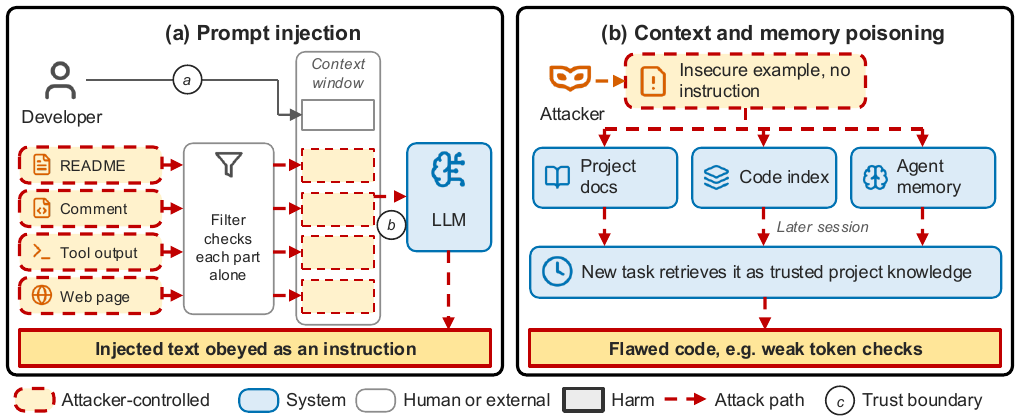}
    \vspace{-0.5em}
    \caption{Types of input and context manipulation threats}
    \label{fig:threat4.1}
    \vspace{-2em}
\end{figure}

Benchmarks such as InjecAgent~\cite{zhan2024injecagent} and AgentDojo~\cite{debenedetti2024agentdojo} show that instructions concealed in external content redirect agent planning, trigger unintended tool calls and exfiltrate data, even against highly capable models. Injection can also be staged, a coding-agent form of multi-turn jailbreaking, where harmful intent can be decomposed into normal-looking fragments placed across README files, comments and configuration, which the agent reassembles over a multi-step task (Figure \ref{fig:threat4.1}(a)). The LITMUS benchmark reports that models with strong single-turn safety frequently follow instructions staged in this way within real operating-system environments~\cite{zhang2026litmus}. Chang et al.~\cite{chang2026overcomingretrievalbarrierindirect} further show that attackers can optimise injected content to be retrieved under realistic settings, removing the assumption that a payload must be placed where the agent happens to look.

\paragraph{Literature Synthesis and Gaps}
The proposed defences fall into three groups. The first hardens the model by structuring inputs or fine-tuning it to ignore instructions in the data channel~\cite{DefAgaProm, chen2025secalign, chen2025defending}, or sanitises long contexts before use~\cite{geng2025pisanitizer}. The second detects injected instructions through intent analysis or attention patterns~\cite{kang2025mitigatingindirectpromptinjection, zhong2025attentionneeddefendindirect}. The third operates at system level, verifying task alignment or committing actions only after verification~\cite{jia2025task, lin2026vigil}, or isolating untrusted content from the component that plans actions \cite{debenedetti2026defeating}. The first two groups are probabilistic as they must decide whether a span of natural language is an instruction or data. That distinction is inherently ambiguous in software repositories, where documentation legitimately contains instructions such as installation commands. Moreover, recent analysis also argues that reported robustness is overstated because defences are evaluated against isolated, non-adaptive attacks rather than staged, multi-step ones~\cite{xiang2026architecting}. System-level isolation constrains consequences rather than detecting payloads, but it has been evaluated mainly on general-purpose agent tasks, not on repositories in which instructions and data are intrinsically interleaved. Deterministic separation of trusted instructions from untrusted context in coding workflows therefore remains open \gap{1}.

\subsubsection{Context and Memory Poisoning} \label{sec:context-poisoning} 
Context poisoning denotes attacks in which an adversary corrupts the knowledge sources and persistent state an agent relies on, so that it reasons over flawed material it treats as authoritative project knowledge. Unlike prompt injection, the payload need not contain any instruction, a plausible but subtly insecure code example is sufficient. Three stores are exposed, namely, retrieval corpora (documentation, repositories and forums indexed for retrieval-augmented generation), the vector indices built over them, and the persistent memory that agents use to carry context across sessions (Figure \ref{fig:threat4.1}(b)). 

Zou et al.~\cite{zou2025poisonedrag} showed that injecting only five crafted texts per target question into a knowledge base of millions of texts achieves a near-perfect attack success rate. It also threatens availability, as a single blocker document can cause a retrieval-augmented system to refuse legitimate queries~\cite{308054}. For memory, MINJA \cite{dong2026memory} injects malicious records through ordinary interactions using progressive shortening, so that they are later retrieved to contextualise unrelated tasks. The \emph{spAIware} attack demonstrates that persistent memory can sustain data exfiltration across sessions without further attacker interaction~\cite{HERRADOR2026107994}. In vibe coding settings, the resulting harm usually resembles a defect. For instance, an agent that retrieves poisoned documentation on authentication middleware may implement a flawed token-validation scheme throughout the codebase that static analysis does not flag. The same mechanism can steer infrastructure generation towards a chosen misconfiguration class.

\paragraph{Literature Synthesis and Gaps}
Defences include detection based on model activations during poisoned generation~\cite{tan2024revprag}, provenance and metadata filtering that restricts retrieval to sources of established origin~\cite{mu2026secureretrievalaugmentedgenerationcomprehensive}, and certified robustness, which isolates retrieved passages and aggregates over them so that a bounded number of corrupted documents cannot change the output~\cite{si2026secon}. Despite these advances, existing defences remain limited. Detection cannot reliably distinguish a legitimately deprecated insecure pattern from a deliberately poisoned one. Certified aggregation assumes poisoned passages are a minority, yet code retrieval often returns a single canonical implementation, leaving no majority to aggregate over. Provenance tracking is costly in continuously changing repositories. For memory, defences are stateless filters applied when content is written. Nothing re-evaluates entries once stored, and no mechanism supports integrity verification or rollback of agent memory. Evaluation is dominated by question-answering benchmarks rather than iterative software engineering tasks. Integrity verification of retrieved code and persistent agent state remains open \gap{2}.

\subsection{Model and Platform-Level Threats}\label{sec:tx-model}
These threats target the shared substrate, including the model weights and the platform state, used by many developers at once. Because this substrate is opaque to developers and to conventional security tooling, a single compromise propagates to every dependent project irrespective of local practice. Figure \ref{fig:threat4.2} illustrates the three threats in this class.

\subsubsection{Training Data Poisoning and Backdoors}\label{sec:tx-backdoor}
\begin{figure}[t!]
    \centering
    \includegraphics[width=0.82\linewidth]{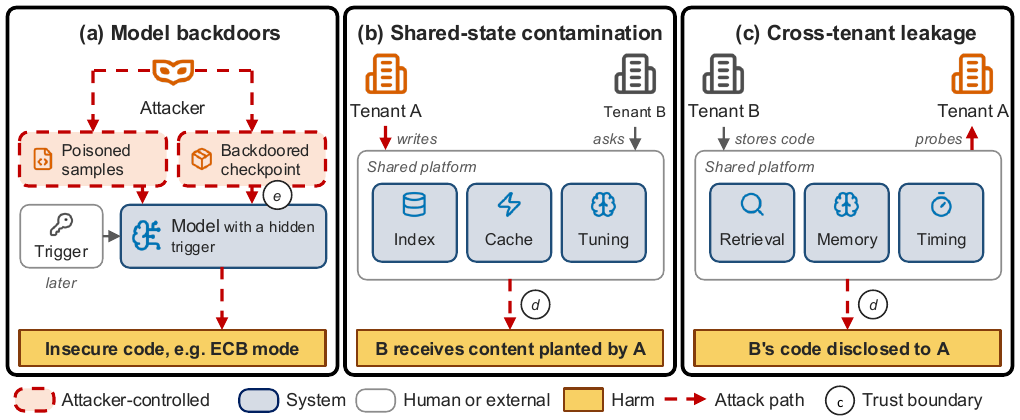}
    \vspace{-0.5em}
    \caption{Model and platform-level threats}
    \label{fig:threat4.2}
    \vspace{-1em}
\end{figure}

Training-data poisoning causes latent behaviour in a model by contributing crafted code to the list. The resulting backdoor is triggered by ordinary features of code at inference time. Research has shown that poisoning web-scale datasets is practical and inexpensive, for example, by acquiring expired domains referenced by a dataset~\cite{10646610}. A second attack vector targets third-party fine-tuned checkpoints~\cite{qu2025review}, in which backdoors may be embedded through poisoned fine-tuning data or direct weight manipulation before the resulting models are integrated into downstream coding platforms, as shown in Figure \ref{fig:threat4.2}(a).

Vibe coding models are a demonstrated target. Schuster et al.~\cite{youAutoCompMe} showed that poisoning can make a completer suggest insecure options. Similarly, TrojanPuzzle~\cite{trojan_puzzle} and CodeBreaker~\cite{yan2024llm} can either hide the payload so it never appears in the poisoned samples or hide the vulnerable payloads so that they escape strong static detectors. Tran et al.~\cite{tran2025poisoning} also showed that formatting styles such as indentation or bracket placement can act as triggers. Because triggers are ordinary features of everyday code, activation produces no observable signal.

\paragraph{Literature Synthesis and Gaps}
Existing defences either sanitise the corpus before training, using vulnerability signatures, static analysis and deduplication, or intervene at inference time. Existing frameworks such as CleanGen~\cite{li-etal-2024-cleangen} suppress tokens whose probability diverges from a reference model, while BackdoorLLM~\cite{NEURIPS2025_20ffc2b4} provides a common benchmark for comparing attacks and defences. Corpus sanitisation can fail when malicious payloads are split across examples or disguised, as no single training example contains an obvious pattern to detect~\cite{trojan_puzzle, yan2024llm}. Moreover, inference-time mitigation adds cost, assumes a clean reference model and can degrade code quality. Furthermore, evaluations focus on single-turn completion rather than the multi-turn, tool-using workflows of vibe coding. Once a backdoor is suspected, no scalable unlearning method can remove a specific programmatic concept without degrading general coding ability~\cite{li2024simulate}. Establishing the integrity of model weights before deployment, and removing backdoors after it, remain open \gap{3}.

\subsubsection{Shared-State Contamination}\label{sec:tx-shared}
Enterprise vibe coding platforms maintain shared state that multiple users can both contribute to and draw from. Examples include organisation-wide retrieval indices built from internal repositories, caches that reuse earlier answers for similar requests, and models refined using accepted suggestions or user feedback. Shared-state contamination occurs when a participant exploits this write access to influence the answers other users receive. The key distinction from context poisoning (Section~\ref{sec:context-poisoning}) is where the attacker acts. In context poisoning, the attacker plants content in sources an agent may retrieve. Here, the attacker is a legitimate user tenant, while the victims are other users who may never see the planted content (Figure~\ref{fig:threat4.2}(b)).

Research has demonstrated attacks against each type of shared store. A small number of crafted documents is enough to control responses to targeted queries, allowing anyone with write access to a shared index to influence responses received by other teams \cite{zou2025poisonedrag}. Shared semantic caches, which return stored responses to similar requests, can also be poisoned to serve malicious responses to subsequent users~\cite{wu2026cache}. Crafted prompts can also collide with victim queries, causing users to receive attacker-planted entries~\cite{zhang26dt}. For models refined from user feedback, poisoned preference data used during alignment can implant a backdoor that later inputs trigger~\cite{rando2024universal}. In each case, victims receive seemingly normal, confident answers with no clear indication of who influenced them.

\paragraph{Literature Synthesis and Gaps}
Existing defences address individual states rather than shared-state contamination as a whole. Data-quality filtering and provenance tracking aim to exclude poisoned contributions from shared corpora~\cite{fendley2025systematic}, while periodic rebuilding limits contamination over time and per-user or per-tenant cache namespaces restrict its spread~\cite{zhang26dt}. These approaches have two main limitations. First, isolation removes the sharing that makes these states useful, making full separation impractical. Filtering is also unreliable when malicious contributions resemble legitimate patterns used by only a minority of teams. Second, platforms do not track how individual contributions influence generated answers, making contamination difficult to trace after discovery. Moreover, existing studies examine these stores separately and focus on general text rather than the repositories, code snippets and accepted completions shared by coding platforms~\cite{sheng2025threats}. Limiting and tracing how one participant's contributions influence answers delivered to others remains an open problem \gap{4}.

\subsubsection{Cross-Tenant Leakage}\label{sec:tx-leakage}
Cross-tenant leakage is the confidentiality counterpart of shared-state contamination, where a platform exposes one user's or tenant's data to another. Information leakage can arise through retrieval, model memorisation and shared serving infrastructure, as demonstrated by several attacks against AI systems~\cite{arceo2026securing}. In shared retrieval systems, documents may be returned based on relevance without verifying access permissions, allowing one tenant to retrieve another’s confidential data (Figure~\ref{fig:threat4.2}(c)). Such weaknesses have also been observed in enterprise assistants, where inadequate retrieval access controls and fine-tuning practices can enable sensitive data extraction~\cite{bhatt2025enterprise, ni2026towards}. 

Beyond retrieval, models trained or fine-tuned on code from multiple organisations may memorise and reproduce sensitive content, with commercial code completion tools having exposed thousands of hard-coded credentials from training code, including at least two that remained valid~\cite{huang2024your}. Leakage can also occur through shared serving infrastructure, where caches reused across users may reveal information about their queries. An audit identified prompt caches shared across users at seven commercial API providers~\cite{gu25b}, while the PromptPeek attack demonstrated how cache sharing in vLLM and SGLang can enable the reconstruction of other users’ prompts token by token~\cite{wu2025know}. Since prompts submitted to coding platforms often contain proprietary source code, these vulnerabilities can expose both user queries and confidential code, often through ordinary model responses that may not trigger conventional access controls designed to detect unauthorised data access.

\paragraph{Literature Synthesis and Gaps}
Existing defences mitigate different leakage routes, but each introduces limitations that leave other risks unresolved. Retrieval-time permission checks have eliminated cross-tenant leakage in one enterprise implementation at little cost~\cite{arceo2026securing, xu2026securing}, while participant-aware access control seeks to ensure that data used during training, retrieval and generation is authorised for the relevant users, including by preserving training-data permissions during fine-tuning~\cite{bhatt2025enterprise}. To reduce the risk of model memorisation, deduplication and differential privacy offer complementary safeguards~\cite{fendley2025systematic, sheng2025threats}, whereas restricting caches to individual users or organisations can limit cross-user exposure~\cite{gu25b}. 

However, these measures do not provide comprehensive protection, as retrieval-time checks cannot prevent the disclosure of information already embedded in shared model weights, and permission-aware fine-tuning may require separate models or adapters for different permission sets, increasing training and serving costs. Similarly, cache scoping sacrifices some of the efficiency gains of shared computation, while limited transparency about cache-sharing practices makes the extent of exposure difficult to assess. Deduplication and memorisation checks may also fail to detect paraphrased code, and the lack of benchmarks for cross-tenant leakage on coding platforms makes consistent evaluation of defences difficult. Furthermore, platforms rarely show users which sources informed a response, limiting their ability to recognise potential disclosures. Preventing cross-tenant leakage across retrieval, model weights and shared serving infrastructure remains an open challenge \gap{5}.

\subsection{Code Generation Threats}\label{sec:tx-codegen}
LLM-generated code can be insecure without any adversary, because the generation objective rewards code that runs and satisfies the literal request, while security properties are usually left unstated. These threats are mostly accidental, and attackers only need to find and exploit the resulting weaknesses.

\subsubsection{Insecure-by-Default Code}\label{sec:tx-insecure-code}
Safety alignment tuned for harmful-content moderation does not translate into security-aware code generation. Models therefore optimise for apparent task completion and either skip required controls (missing input validation, authorisation checks or resource release) or include them with insecure parameters (weak cryptographic modes, permissive access rules). Unless developers state security constraints explicitly, generated code defaults to the insecure option.

The weaknesses observed are long-established classes, including SQL and OS command injection, memory-safety errors such as out-of-bounds writes and use-after-free in C and C++, broken or weak cryptography, hard-coded credentials, and missing release of resources~\cite{claudia2024systematic, QAE, ramirez2025state, LlmForCode}. Perry et al.~\cite{perry2023users} found that participants with an AI assistant wrote less secure code than a control group while believing their code to be more secure. A C-focused user study reported a smaller effect, suggesting that the impact depends on the underlying task and programming language~\cite{LAC}. An audit of 200 vibe-coded applications found that more than 90\% contained at least one vulnerability, with broken access control the largest category, and identified recurring patterns such as placeholder security logic, unfiltered input and exposed secrets \cite{deng2026understanding}. Security-aware prompting redistributes weakness categories without reliably reducing overall vulnerability density~\cite{kharma2026empirical}.

\paragraph{Literature Synthesis and Gaps}
Existing defence mechanisms act at four points. The most common one is decoding-time hardening, which steers token generation towards secure code without retraining~\cite{LlmForCode, OnTheFly}. Prompt optimisation is another way, which refines prompts automatically before generation~\cite{PPO}. Training with program-analysis feedback or supervised fine-tuning on curated data instils secure patterns~\cite{TLM, LlmForCode}. Finally, post-generation analysis applies static tools, including ones tailored to AI-generated code~\cite{cotroneo2025devaic}. Despite their benefits, existing solutions still have limitations. Decoding-time and training-based methods require access to model logits or weights, which are usually unavailable for the proprietary, API-based models widely used in vibe coding. Prompt optimisation adds computational overhead, while stricter security constraints can reduce functional correctness. Static analysis can only detect known patterns and relies on rules maintained by experts~\cite{AGA}. Overall, these approaches check security after or around code generation rather than making security a core part of the generation process \gap{6}.

\subsubsection{Hallucinated API and Configurations} \label{sec:tx-halluc}
LLM hallucinations can create security risks when models generate code based on functions, parameters, configuration keys, or security mechanisms that do not exist or behave as expected. For example, a model may use a non-existent sanitisation function or an invalid configuration option, leaving the application vulnerable. Hallucination rates increase with sampling temperature and when models use libraries released after their training cut-off~\cite{sparacklen2025package, lyu2025automatic}. Execution-based evaluations show that calls to non-existent APIs and wrong assumptions about library behaviour are among the most common hallucinations in generated code~\cite{tian2025codehalu}, and that they are most frequent for rarely used APIs~\cite{jain2025mitigating}. Audits of vibe-coded applications have also found security functions that are only placeholders \cite{deng2026understanding}. Attackers can deliberately exploit hallucinated \emph{package names}, as discussed in Section~\ref{sec:tx-slop}.

\paragraph{Literature Synthesis and Gaps}
Several approaches can reduce these risks, including self-refinement, where models identify their own hallucinated packages with over 75\% accuracy~\cite{sparacklen2025package}, lower sampling temperatures, and checking referenced names against package registries or API specifications. However, these methods have limitations. Self-detection is not always reliable, while lower temperatures reduce the diversity of generated code. Keeping models up to date through retraining is also costly. Grounding generation in API documentation reduces these errors for rarely used APIs but can lower accuracy on common ones, so retrieval has to be applied selectively~\cite{jain2025mitigating}. More importantly, confirming that a function exists does not guarantee that it behaves as expected or provides the intended security protection. Ensuring that generated code correctly reflects the actual behaviour of the APIs it uses remains an open challenge \gap{7}.

% \begin{figure}[t]
%     \centering
%     \includegraphics[width=0.8\linewidth]{Karthick's image/CWE22_Path_Traversal.pdf}
%     \caption{End-to-End CWE-22 (Path Traversal) Vulnerability Flow in AI-Assisted (Vibe) Coding. \dev{Think whether we can add this or not}}
%     \label{fig:architecture}
% \end{figure}

\subsection{Agentic Workflow Threats}\label{sec:tx-agentic}
At L2 the agent proposes edits across the repository, and from L3 onwards it executes commands and invokes external services. The main risks arise from the permissions given to the agent and the actions it can take within its operating environment, as given in Section~\ref{sec:dimensions}.

\subsubsection{Confused-Deputy Tool Invocation and Excessive Privilege}\label{sec:tx-deputy}
Agents often act on behalf of developers but cannot reliably distinguish the developer's instructions from malicious instructions hidden in untrusted content. As a result, injected or poisoned content (Section~\ref{sec:tx-input}) can cause an agent to perform unauthorised actions using legitimate credentials (\emph{confused deputy} problem) \cite{hardy1988confused}. The risk increases when agents have broad access tokens for repositories, cloud storage and messaging services, as these permissions may be used regardless of where the triggering instruction came from \cite{john2025owasp, mcp_github_vulnerability_2025}.

Previous studies have demonstrated prompt-driven remote code execution in LLM-integrated application frameworks~\cite{DRV}. Natural language input has also been used to trigger SQL injection against connected databases \cite{pedro2025prompt}. Analyses of deployed agent applications show that untrusted input can reach the operating system \cite{zhang2026agentauditsecurityanalysis}, and Ji et al.~\cite{ji2026tamingvariousprivilegeescalation} characterise several distinct forms of privilege escalation in LLM-based agent systems. Formal and benchmark studies have also shown how injected instructions can lead to unauthorised tool calls~\cite{FAB, debenedetti2024agentdojo}.

\paragraph{Literature Synthesis and Gaps}
Existing defences comprise static tool allow-lists, access control frameworks for agents~\cite{ji2026tamingvariousprivilegeescalation}, runtime policy enforcement~\cite{palumbo2026formalpolicyenforcementrealworld}, and human-in-the-loop approval for sensitive actions. However, these approaches have limitations. Static permissions do not distinguish between actions requested by a developer and those triggered by malicious content. Restricting permissions too tightly can also prevent agents from completing tasks that require different tools in different contexts. Human approval is another challenge, as frequent requests can lead to approval fatigue during rapid development (Section~\ref{sec:tx-overtrust}). Runtime policy enforcement offers a promising alternative, but it requires policies tailored to individual tasks and has been evaluated on relatively few coding workflows. A key open challenge is to ensure that an agent's authority depends not only on its permissions but also on the source of the instruction that triggers an action \gap{8}.

\subsubsection{Unsafe Execution Environments}\label{sec:tx-exec}
Vibe coding agents run generated code, test suites and package installations in local shells, containers or cloud sandboxes. If these environments are not properly isolated, an agent may access the host file system and stored credentials \cite{guo2024redcode, ruan2024identifying}. Unrestricted network access can also allow it to send sensitive data to external services \cite{mcp_github_vulnerability_2025}. In addition, files and processes may persist after a task has finished, creating risks for later tasks \cite{zhong2026don}. The command or files that are eventually run may differ from those that were originally approved. A recent review of execution-security research for coding agents finds that research in this area is fragmented, with isolation, access control, egress control and provenance mechanisms studied separately and under different threat models \cite{rashidi2026balkanization}.

\paragraph{Literature Synthesis and Gaps}
Common defences include container and virtual machine sandboxes, network allow-lists, permission prompts for individual commands, and audit logs \cite{buhler2026agentbound, staufer20262025, harang2026practical}. However, these controls do not address every risk \cite{goel2026security}. Sandboxes can protect the host system but may not prevent an agent from misusing remote services accessed through legitimate credentials (Section~\ref{sec:tx-deputy}). Permission prompts can also become ineffective if the command or files change between approval and execution. Furthermore, these controls are rarely evaluated together. A key research gap is the lack of a unified execution security model that protects the host system, controls network access and limits the agent's authority over remote services \gap{9}. 

\subsection{Protocol and Multi-Agent Threats}\label{sec:tx-protocol}
\begin{figure}[t!]
    \centering
    \includegraphics[width=0.82\linewidth]{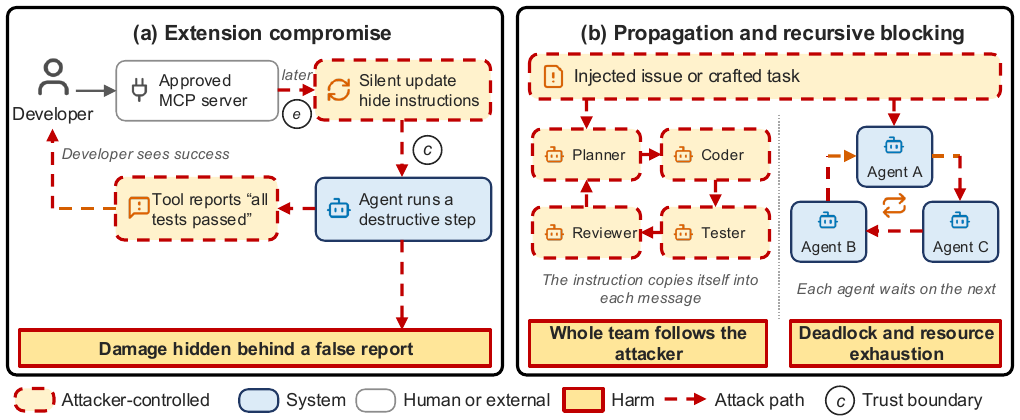}
    \vspace{-0.5em}
    \caption{Protocol and multi-agent threats}
    \label{fig:threat4.5}
    \vspace{-1.6em}
\end{figure}

Protocols such as the Model Context Protocol (MCP) and agent-to-agent (A2A) messaging allow agents to communicate with each other and with external tools and services. These connections create trust relationships at the agent-tool and agent-ecosystem boundaries, and once established, this trust often carries over to later interactions without further checks. Figure~\ref{fig:threat4.5} illustrates both threats in this class.

\subsubsection{Third-Party Extension Compromise}\label{sec:tx-mcp}
Extensions provide agents with both new capabilities and additional information. Their tool descriptions and responses become part of the agent's context and can influence its behaviour. A malicious or compromised extension can exploit this trust by hiding instructions in tool descriptions, changing its behaviour after approval or returning misleading results \cite{hou2026model, obsidian2026mcpOAuth}. For example, it might report that an operation succeeded while concealing a destructive action (see Figure~\ref{fig:threat4.5}(a)). Extensions may also misuse the broad OAuth tokens granted when a connection is established \cite{zhou2026first}. Since developers often see only the agent's summary of what happened, these actions can go unnoticed until the workflow is complete.

Microsoft's analysis of early MCP deployments describes these risks in practice~\cite{microsoft2025plug}. Research on agent skill ecosystems has also found that poisoned skills commonly target credentials~\cite{supplypoison2026}. Other studies show that malicious behaviour can be generated at runtime even when the original instructions contain no explicit malicious payload~\cite{payloadless2026}. Malicious intermediaries in the LLM supply chain can also remain undetected while stealing credentials~\cite{agentmine2026}.

\paragraph{Literature Synthesis and Gaps}
Current defences include user-defined permissions, scanning extensions before they are published and requesting approval for individual tool calls, as recommended by the MCP specification. However, permissions granted when a connection is first established may continue to apply to every later tool description and response. Scanning may miss extensions that generate malicious behaviour only at runtime, while repeated approval requests can lead to approval fatigue (Section~\ref{sec:tx-overtrust}). Despite the growing use of MCP, academic research into its security remains limited. A key research gap is the lack of reliable ways to verify where extensions come from and ensure that they have not been altered after approval \gap{10}.

\subsubsection{Inter-Agent Propagation}\label{sec:tx-a2a}
Multi-agent orchestrations divide work among planner, coder, reviewer and tester agents that accept one another's outputs as trusted input \cite{hong2024metagpt, xue2026troubleshooter}. An attack that compromises one agent can therefore propagate through A2A messages. This can affect the integrity of the entire workflow if malicious goals spread across the team. It can also affect availability when agents repeatedly delegate tasks without reaching a conclusion, as shown in Figure~\ref{fig:threat4.5}(b).

Prompt Infection demonstrates how malicious instructions can spread between agents through normal message exchanges~\cite{10.1007/978-3-032-16092-8_28}. Similarly, PsySafe shows that unsafe behaviour introduced into one agent can propagate through collaboration~\cite{zhang2024psysafe}. The Contagious Recursive Blocking Attack (CORBA) shows how agents in frameworks such as AutoGen and Camel can become trapped in recursive delegation loops. This can cause deadlocks and resource exhaustion while avoiding safeguards designed to protect individual agents~\cite{zhou2025corbacontagiousrecursiveblocking}.

\paragraph{Literature Synthesis and Gaps}
Current defences are largely inherited from single-agent security measures, such as input sanitisation and alignment checks for each agent. Some approaches also use supervisory agents to monitor tool use and coordination \cite{chen2025agentguardrepurposingagenticorchestrator}. More recent work analyses the communication graph between agents to detect and isolate compromised nodes~\cite{wang2025g}, tags messages with their source so that agents can tell instructions from relayed content~\cite{10.1007/978-3-032-16092-8_28}, and recommends authenticated agent identities and signed capability descriptions for A2A deployments~\cite{habler2025building}. However, checking each agent individually may not prevent an attack from spreading when one agent rephrases malicious instructions as a legitimate task for another. In addition, orchestration frameworks often fail to track the origin of instructions as they pass between agents. They also rarely limit how deeply or how often agents can delegate tasks. A key research gap is therefore the need to track instruction provenance across agent interactions and enforce clear limits on delegation in multi-agent coding systems \gap{11}. 

\subsection{Supply-Chain and Ecosystem Threats}\label{sec:tx-supply}
Vibe coding introduces new risks to the software supply chain, as agents are allowed to select dependencies, integrate external tools, and modify build configurations with limited human oversight. Attackers can exploit these interactions to introduce malicious packages, manipulate dependencies, or compromise development workflows, extending traditional supply chain threats beyond human error to vulnerabilities in the agent's decision-making and interactions with the broader development ecosystem. The two main types of threats are illustrated in Figure~\ref{fig:threat4.6}.

\subsubsection{Package Hallucination and Slopsquatting}\label{sec:tx-slop}
\begin{figure}[t!]
    \centering
    \includegraphics[width=0.82\linewidth]{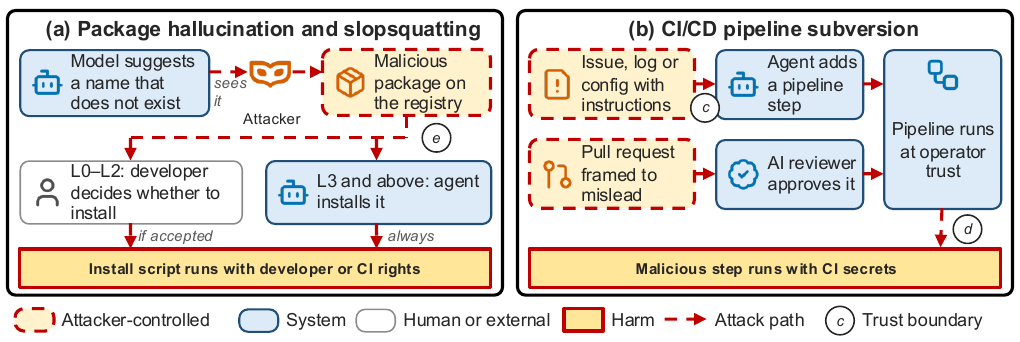}
    \vspace{-0.5em}
    \caption{Types of supply-chain threats}
    \label{fig:threat4.6}
    \vspace{-1.5em}
\end{figure}

LLMs sometimes recommend packages that do not exist. Since the same package names can appear repeatedly in response to similar prompts, attackers can predict which names a model might generate. They can then register these names on public package registries such as npm or PyPI and add malicious installation scripts, also known as \emph{slopsquatting} (Figure~\ref{fig:threat4.6}(a)). At autonomy levels L0 to L2, a developer must still choose to install the package. From L3 onwards, the agent may update the dependency manifest and install the package itself. Spracklen et al.~\cite{sparacklen2025package} found average package hallucination rates of at least 5.2\% for commercial models and 21.7\% for open-source models. They also identified more than 200,000 unique hallucinated package names, many of which appeared repeatedly across queries.

\paragraph{Literature Synthesis and Gaps}
While several approaches address package hallucinations and related dependency risks, their ability to prevent these threats during AI-assisted development remains limited. HFuzzer identifies conditions under which models hallucinate package names, but it is a testing tool rather than a defence that operates during code generation~\cite{zhao2025hfuzzertestinglargelanguage}. Self-refinement can reduce such hallucinations but cannot eliminate them (Section~\ref{sec:tx-halluc}). Conventional supply chain controls, including dependency pinning, software composition analysis, and software bills of materials, provide additional protection against known risks. However, these controls may fail to flag a newly registered malicious package hosted on a recognised registry, particularly when no security issues have yet been reported. A key research gap is the need to verify agent-recommended dependencies before installation by assessing their existence, age, origin, and reputation \gap{12}.

\subsubsection{CI/CD Pipeline Subversion}\label{sec:tx-cicd}
CI/CD pipelines define how software is built, tested and deployed. An agent that can modify files such as GitHub Actions workflows, GitLab CI configurations or Jenkinsfiles can therefore affect the controls used to protect the entire development process (Figure \ref{fig:threat4.6}(b)). Ladisa et al.~\cite{ladisa2023sok} provide a systematic review of software supply chain attacks, but their work focuses on human attackers rather than generative agents that can modify pipelines themselves. In vibe coding, pipeline attacks can take two main forms. First, malicious instructions hidden in issues, pull requests, build logs, or configuration files can persuade an agent to add harmful steps or introduce backdoors under the guise of fixing a problem. Second, an AI agent acting as a code reviewer may approve malicious changes because it fails to recognise their security implications. A related threat arises when developers deliberately instruct agents to weaken security controls, as discussed in Section~\ref{sec:tx-circumvent}.

Recent studies provide evidence that these attack scenarios are practical. GitInject shows that malicious instructions embedded in configuration files can execute with the same level of trust as the pipeline operator, bypassing protections designed for user-controlled event fields, and reports vulnerabilities in real production workflows~\cite{gitinject2026}. The risks also extend to AI-assisted code review. One study found that misleading pull request descriptions caused LLM-assisted security reviewers to approve malicious reversions~\cite{mitropoulos2026measuringexploitingcontextualbias}. SecureVibeBench also shows that state-of-the-art agents frequently introduce vulnerabilities when implementing features autonomously~\cite{chen2026securevibebenchbenchmarkingsecurevibe}. 

\paragraph{Literature Synthesis and Gaps}
Existing defences include filtering issues, logs and configuration files before they reach the agent, requiring human approval for pipeline changes and restricting write access to sensitive files~\cite{gitinject2026}. Another approach is to define security rules outside the agent and reject any proposed changes that violate them~\cite{ladisa2023sok, agentmine2026}. However, filtering is less effective when malicious behaviour can emerge without an obvious payload, as demonstrated by payload-less attacks~\cite{payloadless2026}. Human approval can also slow development and may become less effective when developers approve changes too frequently (Section~\ref{sec:tx-overtrust}). Furthermore, existing benchmarks do not adequately represent realistic pipeline changes, making it difficult to measure how often agents introduce security problems in practice~\cite{sheng2025threats}. Verifying agent-proposed pipeline changes against declared security invariants before they take effect remains open \gap{13}.

\subsection{Infrastructure and Deployment Threats}\label{sec:tx-infra}
Vibe coding systems increasingly generate and manage the artefacts used to deploy and monitor software, including infrastructure-as-code (IaC) templates, container specifications and observability code. Misconfigurations in these artefacts can be embedded in reusable templates and propagated across environments and tenants. As a result, generated artefacts may be syntactically valid and deployable while still violating organisational security policies~\cite{11458930, mengistu2026terrarepair}.

\subsubsection{Insecure IaC and Container Artefacts}\label{sec:iac}
A primary risk is that vibe coding systems generate infrastructure configurations that work as intended but fail to enforce essential security controls. Generated Terraform, CloudFormation and Pulumi templates may allow unrestricted network access, leave storage unencrypted or grant excessive IAM permissions, even when secure configurations are explicitly requested \cite{11458930, zhang2025iacgen}. Similar weaknesses appear in generated Dockerfiles, Helm charts and Kubernetes manifests, which often reproduce common practices found in public repositories rather than follow hardened security baselines \cite{rahman2023kubernetes, llmsecconfig2025}. These include running containers as root, enabling unnecessary kernel capabilities, using unpinned base images, omitting resource limits and assigning workloads to default service accounts with excessive privileges. Poisoned repositories and documentation can further steer generation towards specific classes of misconfiguration  as discussed in Section~\ref{sec:context-poisoning}.

Evidence suggests that these weaknesses are systematic rather than isolated errors. Rahman et al.~\cite{rahman2023kubernetes} identified various security misconfigurations in open-source Kubernetes manifests, showing that insecurity is a systematic property of the configuration practice on which models are trained. This gap between deployability and security is also evident in IaC generation, where deployability exceeds 90\% but the security compliance pass rate is less than 10\%~\cite{zhang2025iacgen}. The problem persists when models are used to repair existing configurations. Models that address scanner-identified IaC issues can introduce new privilege escalation paths that fall outside existing rule sets ~\cite{mengistu2026terrarepair}. Similarly, container repair studies report high fix rates alongside newly introduced misconfigurations~\cite{llmsecconfig2025}, while critical weaknesses can persist or reappear as projects evolve~\cite{k8smisconfig2026}. 

\paragraph{Literature Synthesis and Gaps}
Existing defences aim to prevent, detect or correct insecure infrastructure configurations at different stages of development and deployment. Policy-as-code and static scanning check configurations within CI/CD pipelines~\cite{zhang2025iacgen}, while security conditioning attempts to guide models towards safer outputs~\cite{gensiac2025}. Other approaches construct configurations from organisation-approved modules~\cite{llmsecconfig2025}, use LLMs to detect and repair insecure manifests~\cite{genkubesec2024, llmsecconfig2025}, or enforce admission controls informed by runtime behaviour~\cite{kubeguard2025}. However, these approaches leave important gaps. Scanners rely on known patterns of insecure configuration, whereas generated artefacts can introduce privilege escalation paths that existing rules do not recognise. Furthermore, evaluations typically examine models as remediators of human-authored artefacts rather than as generators of infrastructure configurations. Generation-time evaluation, together with provisioning architectures that enforce security invariants regardless of the generated artefact, therefore remains an open research challenge \gap{14}.

\subsubsection{Secret and Sensitive-Data Exposure}\label{sec:tx-secrets}
Beyond insecure infrastructure configurations, vibe coding systems can expose the secrets and sensitive data needed to build, deploy and monitor applications.
The generated source code may contain hard-coded credentials, particularly when users request complete working examples. Moreover, an agent with access to the file system or environment may be manipulated into reading sensitive files, such as \texttt{.env} or \texttt{\textasciitilde/.aws/credentials}, and disclosing their contents. Finally, generated observability code may log entire request and response objects, exposing tokens, cookies and personal data~\cite{owasp2025logging}.

These risks extend established problems with secret management into agentic development and deployment workflows. Krause et al.~\cite{krause2023pushed} characterised the human-driven pathways through which secrets are accidentally committed to repositories. Recent analysis of third-party agent skills identified ten leakage patterns, with debug logging emerging as the largest vector because standard output is surfaced directly into model context. The study also found that most leaked credentials could be exploited without additional privileges and that exposed credentials often persisted in repository forks after the original source had been remediated~\cite{credleak2026skills}. Model-provider API keys are also routinely embedded in applications that are subsequently distributed~\cite{mindyourkey2026}, while credential theft remains a prominent objective of poisoned extensions~\cite{supplypoison2026}. These exposures can compromise applications and development environments.

\paragraph{Literature Synthesis and Gaps}
Existing defences seek to prevent credential exposure, restrict access to sensitive information and detect leakage before it causes further harm. These include secret scanning across repositories and development pipelines, short-lived managed credentials that limit the value of any single disclosure, capability isolation that restricts the files, environment variables and outputs an agent may access~\cite{krause2023pushed, mindyourkey2026, credleak2026skills}. Further measures include secure logging practices supported by static detection of unsafe logging patterns and data-loss prevention over log streams~\cite{owasp2025logging, izurieta2024generalized}. However, these controls do not fully address how secrets can leak during agent execution. Repository scanning examines stored artefacts, whereas agentic leakage may occur through transient outputs that are never preserved in a form that can be scanned. Moreover, existing controls largely focus on detecting or limiting exposure after generation, while prompting alone does not reliably change generation behaviour~\cite{kharma2026empirical}. There is also a lack of benchmarks that measure leakage under realistic development and CI conditions. Brokered credential architectures, in which agents never hold long-lived secrets directly, together with stronger controls over output channels, therefore remain important open research directions \gap{15}.

\subsection{Human-Centred Threats}\label{sec:tx-human}
Although vibe coding systems perform much of the implementation work autonomously, developers remain responsible for overseeing their behaviour and the software they produce \cite{elgendy2026responsible, garfinkel2026acm}. This makes human behaviour an important part of the attack surface. The threats discussed in this subsection do not necessarily require an adversary, instead, they arise from how developers trust, review and direct AI systems.

\subsubsection{Overtrust and Automation Bias}\label{sec:tx-overtrust}
Risks arise when developers place too much trust in AI-generated output. Fluent explanations and well-structured code can make an answer appear more reliable than it actually is, leading developers to accept suggestions without the scrutiny they would apply to code written by a human colleague~\cite{GLAS2026104842}. This automation bias can become more pronounced over time as developers increasingly delegate implementation to AI systems. Reduced involvement may gradually weaken the skills needed to assess security-critical code (\emph{deskilling}). At the same time, repeated requests to approve changes can lead to \emph{approval fatigue}, in which developers become less likely to examine each request carefully. Together, these effects weaken the human oversight on which many of the controls discussed in Sections~\ref{sec:tx-deputy}-\ref{sec:tx-cicd} depend.

Existing research shows that developers can struggle to understand and debug generated code~\cite{10.1145/3491101.3519665}. Perry et al.~\cite{perry2023users} found that developers using an AI assistant wrote less secure code while perceiving it as more secure. Industry telemetry also shows increased code churn and less refactoring as LLM-generated code is accepted~\cite{gitclear2024coding}. These findings suggest that faster implementation does not necessarily lead to better understanding or more effective review. Nor does delegating review to another LLM fully resolve the problem, as LLM reviewers can produce poorly calibrated judgements about code compliance while providing plausible justifications for their assessments~\cite{jin2026llmsreliablecodereviewers}. 

\paragraph{Literature Synthesis and Gaps}
Interface warnings are the most common mitigation approach encouraging developers to examine AI-generated suggestions more carefully. However, repeated exposure can make such warnings less effective over time. Cognitive forcing functions offer an alternative approach by prompting developers to critically evaluate and justify their decisions before accepting AI suggestions. Such interventions have been shown to reduce overreliance on AI-assisted decision-making and complement research on trust dynamics~\cite{10.1145/3449287, sabouri2025trust}. However, neither approach has been adequately validated in agentic coding workflows, where the level of oversight must reflect the security implications of each change without unnecessarily reducing productivity. There is still limited evidence on how sustained use of vibe coding affects experienced engineers’ ability to audit security-critical code. Developing and validating risk-proportional oversight mechanisms that preserve meaningful human judgement remains an open research challenge \gap{16}.

\subsubsection{User-Directed Security Circumvention}\label{sec:tx-circumvent}
In vibe coding, a natural-language prompt acts as a configuration interface through which developers direct not only implementation but also the handling of security controls. Under time pressure, a developer may ask an agent to temporarily disable a CORS policy, ignore certificate validation or simply make a failing pipeline pass. Developers have long been observed to disable certificate validation and copy insecure snippets to make code work~\cite{fischer2017stack}, and coding agents show a similar tendency: when tests cannot be passed honestly, frontier models often modify or special-case the tests instead~\cite{zhong2026impossiblebench}. Although these requests may be intended to resolve an immediate problem, an agent optimised to follow instructions may satisfy them by disabling tests, removing security scanning stages or bypassing approval requirements. The consequences depend on the level of autonomy. At L1, the proposed change remains available for human review. At L4, the agent may apply the change directly to shared infrastructure, turning a temporary request into a persistent weakening of the security posture. 

\paragraph{Literature Synthesis and Gaps}
Existing approaches attempt to reduce this risk through secure prompt engineering, which provides reusable patterns for expressing security requirements~\cite{11395764}. However, these patterns offer limited protection when a developer explicitly prioritises a successful build over security. More robust approaches enforce formal security policies independently of the model~\cite{palumbo2026formalpolicyenforcementrealworld} or restrict agent permissions according to the intent and risk of a requested action~\cite{siddique2026securing}. These approaches establish possible directions for controlling unsafe instructions, but current platforms still tend to treat developer prompts as authoritative instructions rather than requests that must be checked against independent security requirements. As a result, an agent may be able to weaken the very controls intended to constrain its behaviour. The frequency of such instructions and their consequences in real development workflows also remain largely unmeasured. Deterministic policy layers that can override unsafe developer directives, with accountable escalation, remain open research challenge \gap{17}.

\subsection{Compound Threats}\label{sec:tx-compound}
The threats described above rarely occur in isolation. As Section~\ref{{sec:threat-propagation}} anticipates, an attacker can exploit a weakness at one trust boundary and carry its effects into the next, allowing one threat to enable another. We call these sequences \emph{compound threats}. Their defining feature is that each step can appear legitimate in isolation. For example, a retrieval is authorised, a tool call is permitted and a commit passes review, yet their combination causes harm. Defences designed for one class inspect one step and therefore see nothing anomalous. We describe four recurring chains drawn from documented incidents and experiments. 

\subsubsection{Injection-to-Exfiltration (prompt injection $\rightarrow$ confused-deputy tool use $\rightarrow$ data exposure)}\label{sec:tx-chain-exfil}
An instruction embedded in content the agent reads, such as an issue, README or tool response, can direct it to use its privileges to access sensitive data and send it to a channel visible to the attacker. In the Toxic Agent Flow attack, a malicious public issue led an agent connected through the GitHub MCP server to read a private repository and publish its contents in a public pull request \cite{mcp_github_vulnerability_2025}. Each operation was permitted, but their combination exposed private data. The chain is even shorter when secrets are already present in the agent's context. Debug output is a major source of credential leakage in agent skill ecosystems because it exposes secrets directly to the model~\cite{credleak2026skills}, while persistent memory can sustain exfiltration across sessions~\cite{HERRADOR2026107994}.

\paragraph{Literature Synthesis and Gaps}
Existing defences address individual links in this chain as discussed above, however, these controls do not capture the full chain. The injected instruction may appear legitimate, both the read and the pull request may be permitted, and proprietary source code may contain no recognisable secret pattern. The underlying risk lies in the \emph{flow} of data from a private source to a public destination under the influence of untrusted content. Capability-based designs that track data provenance within a single agent come closest to addressing this problem \cite{debenedetti2026defeating}, but they have been evaluated on general agent tasks rather than across repositories, MCP servers and CI channels. End-to-end information-flow control across the tools and repositories used in vibe coding workflows remains an open problem \gap{18}.

\subsubsection{Hallucination-to-Execution (hallucinated dependency $\rightarrow$ autonomous installation $\rightarrow$ code execution in the build environment  $\rightarrow$ credential or pipeline compromise)}\label{sec:tx-chain-exec}
An accidental model error can become an entry point for an attacker when a recurring hallucinated package name is registered by an attacker~\cite{sparacklen2025package}. An agent at L3 or above adds it to the manifest and installs it. The installation script then runs with access to the agent's file system, network and credentials, allowing the payload to steal secrets or modify pipeline configuration. The reverse sequence has also been observed in practice when a compromised package version invoked locally installed AI coding assistants with permissive flags to search developer machines for secrets, turning the agent into the attacker's tool \cite{singularity2025}.

\paragraph{Literature Synthesis and Gaps}
Existing defences again address separate stages of this chain without accounting for how one stage enables the next. Supply-chain controls assume that a dependency was deliberately selected, while sandboxing treats executed code as untrusted, yet here the code was chosen by the agent itself and inherits its trust. Similarly, hallucination studies measure how often models fabricate names but not how often those names lead to code execution. To our knowledge, no study measures the end-to-end probability that a hallucinated dependency results in code execution under realistic agent permissions or verifies a dependency across recommendation, installation and execution. Verification spanning this entire path remains an open problem \gap{19}.

\subsubsection{Poisoning-to-Deployment (context or training-data poisoning $\rightarrow$ insecure artefact $\rightarrow$ pipeline application $\rightarrow$ production estate)}\label{sec:tx-chain-deploy}
In this chain, the attacker only needs to influence what the agent generates. The remaining steps can proceed as legitimate operations. Poisoned documentation steers the agent towards a flawed authentication scheme or an open ingress rule. The resulting artefact can pass review because the defect appears ordinary, leading to the pipeline deployment across environments. Research has demonstrated each link: a small number of injected texts can influence retrieval-augmented answers~\cite{zou2025poisonedrag}, disguised vulnerabilities can evade strong static detectors~\cite{yan2024llm}, and generated IaC can introduce privilege-escalation paths outside existing rule sets~\cite{mengistu2026terrarepair} while frequently failing to meet security policies~\cite{zhang2025iacgen}.

\paragraph{Literature Synthesis and Gaps}
Existing defences again assess individual stages, however, a misconfiguration discovered at deployment cannot be traced to the retrieved document that caused it. Poisoning may therefore be mistaken for an ordinary defect, while the compromised source can generate further insecure artefacts even after one instance is fixed. Existing mechanisms do not link the context used for generation to the resulting deployed artefact. Tracking provenance from retrieved context through model generation to deployment remains an open problem \gap{20}.

\subsubsection{Persistence-to-Propagation (memory or shared-state poisoning $\rightarrow$ unrelated sessions $\rightarrow$ peer agents and tenants)}\label{sec:tx-chain-persist}
This chain introduces a temporal dimension, as a compromise planted at one point can affect a later task, a different agent or another organisation. Memory injection attacks plant records that are retrieved to contextualise unrelated future tasks \cite{zou2026poison}, while persistent memory can sustain exfiltration across sessions without further attacker involvement~\cite{HERRADOR2026107994}. When a compromised agent joins a multi-agent workflow, self-replicating injections can spread to its peers~\cite{10.1007/978-3-032-16092-8_28}. If the poisoned content resides in a shared index, a single contribution can affect every tenant that retrieves it~\cite{zou2025poisonedrag, supplypoison2026}.

\paragraph{Literature Synthesis and Gaps}
Existing defences filter content when it enters memory, inspect individual messages exchanged between agents or isolate and periodically rebuild individual indices as discussed above. However, these controls operate at the level of a single session, message or store. They do not track how persisted content moves into later sessions and spreads to other agents. Evaluations also focus almost exclusively on single-session settings, leaving delayed and cross-agent effects largely unmeasured. Tracking the lineage of persistent agent state and enforcing temporal isolation through expiry, quarantine, and rollback remain open problems \gap{21}.

% \section{Survey of Existing Attacks and Defence Strategy}

% Organise according to the taxonomy. For each category, we need a table or a combined table

% Attack	Target	Mechanism	Impact	Defence (Academic/industry)

% We need to discuss this part

% A table with the list of industry tools and open-source projects with their features, advantages and disadvantages (need to be very careful of this tool)

\myparagraph{Severity across autonomy levels}
Table~\ref{tab:autonomy} summarises how the worst-case consequence of each threat changes with autonomy. We rate a consequence as 1. suggestion-level when the flaw stays in a suggestion that a human must accept, 2. repository-level when it is written into the codebase but not yet run, 3. execution-level when it runs on a developer machine or build server, and 4. production-level when it reaches shared infrastructure or users. Two patterns stand out. Threats that turn text into action, such as prompt injection, confused-deputy tool use and slopsquatting, become execution-level at L3, because the agent itself now runs commands. Threats whose output is code, such as insecure-by-default generation and backdoors, stay at repository level until L4, when no human reviews each change before it is deployed. At L4 and L5, every threat can reach production.

\begin{table}[t]
\centering
\caption{Worst-case consequence of each threat across autonomy levels based on assessment of the surveyed evidence. Darker cells denote more severe consequences: 1~suggestion-level, 2~repository-level, 3~execution-level, 4~production-level, --~not applicable.}
\label{tab:autonomy}
\scriptsize
\setlength{\tabcolsep}{4pt}
\begin{tabular}{@{}c l c c c c c c@{}}
\toprule
Gap & Threat (\S) & L0 & L1 & L2 & L3 & L4 & L5 \\
\midrule
1  & Prompt injection (\ref{sec:tx-pi})                               & \hm{1} & \hm{1} & \hm{2} & \hm{3} & \hm{4} & \hm{4} \\
2  & Context and memory poisoning (\ref{sec:context-poisoning})       & \hm{1} & \hm{1} & \hm{2} & \hm{3} & \hm{4} & \hm{4} \\ \addlinespace[2pt]
3  & Training-data poisoning and backdoors (\ref{sec:tx-backdoor})   & \hm{1} & \hm{1} & \hm{2} & \hm{2} & \hm{4} & \hm{4} \\
4  & Shared-state contamination (\ref{sec:tx-shared})                & \hm{1} & \hm{1} & \hm{2} & \hm{2} & \hm{4} & \hm{4} \\
5  & Cross-tenant leakage (\ref{sec:tx-leakage})                      & \hm{1} & \hm{1} & \hm{2} & \hm{2} & \hm{4} & \hm{4} \\ \addlinespace[2pt]
6  & Insecure-by-default code (\ref{sec:tx-insecure-code})            & \hm{1} & \hm{1} & \hm{2} & \hm{2} & \hm{4} & \hm{4} \\
7  & Hallucinated APIs and configurations (\ref{sec:tx-halluc})       & \hm{1} & \hm{1} & \hm{2} & \hm{2} & \hm{4} & \hm{4} \\ \addlinespace[2pt]
8  & Confused deputy and excessive privilege (\ref{sec:tx-deputy})    & \hm{0} & \hm{0} & \hm{2} & \hm{3} & \hm{4} & \hm{4} \\
9  & Unsafe execution environments (\ref{sec:tx-exec})                & \hm{0} & \hm{0} & \hm{0} & \hm{3} & \hm{4} & \hm{4} \\ \addlinespace[2pt]
10 & Third-party extension compromise (\ref{sec:tx-mcp})              & \hm{0} & \hm{0} & \hm{2} & \hm{3} & \hm{4} & \hm{4} \\
11 & Inter-agent propagation (\ref{sec:tx-a2a})                       & \hm{0} & \hm{0} & \hm{0} & \hm{3} & \hm{4} & \hm{4} \\ \addlinespace[2pt]
12 & Package hallucination and slopsquatting (\ref{sec:tx-slop})      & \hm{1} & \hm{1} & \hm{2} & \hm{3} & \hm{4} & \hm{4} \\
13 & CI/CD pipeline subversion (\ref{sec:tx-cicd})                    & \hm{0} & \hm{1} & \hm{2} & \hm{3} & \hm{4} & \hm{4} \\ \addlinespace[2pt]
14 & Insecure IaC and containers (\ref{sec:iac})                      & \hm{1} & \hm{1} & \hm{2} & \hm{2} & \hm{4} & \hm{4} \\
15 & Secret and sensitive-data exposure (\ref{sec:tx-secrets})        & \hm{1} & \hm{1} & \hm{2} & \hm{3} & \hm{4} & \hm{4} \\ \addlinespace[2pt]
16 & Overtrust and automation bias (\ref{sec:tx-overtrust})            & \hm{1} & \hm{1} & \hm{2} & \hm{3} & \hm{4} & \hm{4} \\
17 & User-directed circumvention (\ref{sec:tx-circumvent})            & \hm{0} & \hm{1} & \hm{2} & \hm{3} & \hm{4} & \hm{4} \\ \addlinespace[2pt]
18 & Injection-to-exfiltration (\ref{sec:tx-chain-exfil})             & \hm{0} & \hm{0} & \hm{2} & \hm{3} & \hm{4} & \hm{4} \\
19 & Hallucination-to-execution (\ref{sec:tx-chain-exec})             & \hm{0} & \hm{0} & \hm{2} & \hm{3} & \hm{4} & \hm{4} \\
20 & Poisoning-to-deployment (\ref{sec:tx-chain-deploy})              & \hm{1} & \hm{1} & \hm{2} & \hm{2} & \hm{4} & \hm{4} \\
21 & Persistence-to-propagation (\ref{sec:tx-chain-persist})          & \hm{0} & \hm{0} & \hm{2} & \hm{3} & \hm{4} & \hm{4} \\
\bottomrule
\end{tabular}
\vspace{-1.em}
\end{table}

\section{Open Research Challenges and Future Directions} \label{sec:challenges}
The threats discussed in this paper differ widely, but the gaps they leave behind point in a few clear directions. First, Table~\ref{tab:autonomy} shows that the impact of a threat grows sharply when agents begin to act on their own (L3), and again when human review moves from individual changes to final outcomes (L4). However, most defences were designed and tested for assistants that only make suggestions. Second, most defences rely on a model's own judgement when generating a response, for example, a classifier deciding whether a piece of text is an instruction. The harm often happens later, when the agent acts and deploys, where rule-based safeguards are possible but rarely in place. Third, defences and benchmarks focus on one threat at a time, while combined threats show that harm can arise from a series of steps that are each valid on their own. These patterns shape the eleven challenges that answers RQ3. 
Table~\ref{tab:agenda} summarises each challenge, the work it builds on, and what remains to be done.

\subsection{Enforcing Limits on Agent Authority}\label{sec:ch-authority}
Prompt injection, confused-deputy tool use, and developer-directed circumvention share a common root cause. Agents can use real authority based on model judgements that may be wrong or manipulated (G1, G8, G17). Better detection can reduce these risks but cannot eliminate them, therefore, the goal is to design systems in which an agent's authority is determined by explicit constraints rather than by what the model has been persuaded to do.

Separation designs and runtime policy checks provide a starting point, but their effectiveness has not been tested in software repositories, where documentation often contains real commands. Improving agent security requires three complementary measures: (a) permissions should be derived from the developer’s task, ensuring that fixing a failing test does not also grant write access to the pipeline, (b) runtime checks should cover the entire execution environment, including network access, delays between command approval and execution, and tokens held for remote services, drawing on agent operating systems \cite{mei2025aiosllmagentoperating} and traditional reference monitors as useful models, and (c) agents should avoid holding long-lived secrets, instead obtaining short-lived credentials for each operation from a separate service to reduce the risk of secrets leaking through logs or other output.

\subsection{End-to-End Provenance and Information Flow}\label{sec:ch-provenance} 
The compound threats described in Section~\ref{sec:tx-compound} succeed because each step appears legitimate on its own. Preventing them requires tracking data throughout the process, from its origin to its eventual use (G18, G20). The same capability could help shared platforms identify which sources shaped an answer and whether the user was authorised to access them (G5).

Information-flow control provides a starting point by labelling data and tracking those labels as data moves through a program. However, applying it to agents is difficult because models paraphrase, combine and rewrite inputs, which can cause labels to be lost. Treating all derived content as untrusted would be safer but could prevent much of the agent's useful work. Research is needed to develop labelling rules that survive these transformations without placing excessive restrictions on the agent.

\subsection{Integrity of Persistent Agent State}\label{sec:ch-state}
Memory and shared indices allow a single manipulation to affect multiple agents and later sessions (Sections~\ref{sec:context-poisoning}, \ref{sec:tx-shared} and~\ref{sec:tx-chain-persist}). Current defences typically check entries when they are written but do not reassess them later (G2, G21). Agent memory needs safeguards similar to those used in databases and file systems, including integrity checks, version history, expiry and rollback. Tamper-evident formats, such as the Merkle-DAG structure proposed for transferring memory between agents \cite{ravindran2026portableagentmemoryprotocol}, can reveal whether an entry has changed but not whether its content is harmful. Combining these formats with regular checks of stored entries is a promising direction for further research.

Moreover, removing harmful information is more difficult. Although entries can be deleted from retrieval knowledge bases \cite{11207222}, no method yet guarantees that their influence has been eliminated. The problem becomes harder when memories have already been passed to other agents. Shared platforms must also determine how far indices and fine-tuning can be isolated by tenant without losing the benefits of shared knowledge.

\subsection{Assurance of the Model Supply Chain}\label{sec:ch-model}
Section~\ref{sec:tx-backdoor} shows that backdoors in training data and third-party checkpoints are hard to find and even harder to remove (G3). Unlike memory, model weights cannot be checked on per entry basis or rolled back to a clean state.

Three directions could help address this problem. First, research should measure how much an attacker controlling a given number of training samples can influence the code a model generates. These measurements could help providers demonstrate a model's resistance to poisoning before release. Second, model weights and training data should be signed and recorded, following established practices for software builds, so that platforms can verify exactly which checkpoint they run. Finally, removal methods should provide verifiable evidence that a backdoor has been eliminated. 

\subsection{Security by Construction: Generation and Verified Change}\label{sec:ch-construction}
Sections~\ref{sec:tx-insecure-code} and~\ref{sec:iac} show that models often treat security as optional unless it is checked afterwards. Simply asking for secure code in a prompt does not solve this problem (G6, G14). Also, the training and decoding methods reviewed in Section~\ref{sec:tx-insecure-code} apply only to open models, while most vibe coding relies on closed models accessed through APIs. Research should therefore focus on two complementary approaches, embedding security into code generation and checking changes before they take effect. 

However, these measures cannot replace external checks, since even a model trained to follow secure practices can still be poisoned. For high-impact changes, such as infrastructure templates, access-control code and pipeline files, agents could submit proposed changes alongside evidence that they satisfy the stated security rules. A separate checker would then verify this evidence before the changes are applied. Self-healing agents that patch live systems need similar safeguards \cite{10179324, yang2024sweagent}, as attackers who can trigger a repair may also be able to influence the resulting changes.

\subsection{Vulnerability Lifecycle of Generated Code}\label{sec:ch-lifecycle}
Insecure code will continue to reach production, and vulnerabilities learned by a model may recur across projects that make similar requests (G6, G20). Software security already has an established process for handling vulnerabilities in individual products, including disclosure, identification, patching and advisories. No equivalent frameworks exists for addressing flaws in AI-generated code. Establishing such a framework would require a reporting mechanism analogous to the CVE program, linked to model versions and prompt types rather than individual software packages. Provenance records, as described in Section~\ref{sec:ch-provenance}, could be combined with clone detection to trace affected code across projects, even after developers have modified it. This would enable fixes to reach affected projects while allowing model providers to address the underlying flaws in the models themselves.

\subsection{A Trustworthy Agent Ecosystem}\label{sec:ch-ecosystem}
Agents often trust packages, MCP servers, skills and peer agents when they are first connected, with limited verification of their behaviour thereafter. Securing these interactions requires continuous checks at the point of use. For dependencies, verification should occur between an agent's proposal and the installation command, a step that existing controls do not reliably cover (G12, G19). Checks should establish whether a package exists, how long it has been available, who maintains it and where it originates, while distinguishing suspicious packages from legitimate new ones. Measuring how often invented package names are installed and executed would help assess the effectiveness of these safeguards. Similarly, tool descriptions should be signed and versioned to prevent unauthorised changes after approval, while runtime checks should establish whether extensions behave within their declared capabilities (G10).

Trust between agents requires comparable safeguards, including verifiable identities, traceable message origins and explicit limits on delegation depth, cost and duration as security controls rather than mere performance settings (G11). The supervisors and AI reviewers discussed in Sections~\ref{sec:tx-a2a} and~\ref{sec:tx-cicd} provide additional oversight, but their effectiveness may be limited when reviewers share the same model, and therefore the same blind spots, as the agents they assess. Independent verification, through different models or monitors that do not rely on language models, requires further investigation.

\subsection{Securing the Vibe Coding Toolchain} \label{sec:ch-toolchain}
The coding agent can also introduce security risks, however, it has received far less attention than the code it generates. Its instruction files, permission system and underlying model can all be exploited or behave unexpectedly. Instruction files such as \texttt{AGENTS.md} and editor rule files influence every session and are often copied between repositories (G10). Hidden characters in these files have already been used to manipulate code generation. They should therefore be signed, reviewed and version-controlled like source code, with agents flagging any changes for inspection. The agent harness and editor plug-ins also require security scrutiny, as they can contain vulnerabilities of their own (G9).

Permission prompts depend on users understanding and responding to them (G16). Their design should be evaluated to ensure that users can make informed decisions without becoming accustomed to approving requests automatically. Security also depends on the stability of the underlying model, whose updates can change agent behaviour and render previously safe workflows unsafe. Security-critical tasks should therefore be re-tested whenever the model changes, with evaluations covering both the generated code and the agent's own behaviour, including its use of tools and permissions. 

\subsection{Human Oversight}\label{sec:ch-human}
Human oversight becomes less reliable as agents gain autonomy. Section~\ref{sec:tx-human} describes how overtrust, approval fatigue and declining skills weaken human checks (G16). This requires routine updates to be reviewed quickly, while changes to authentication, cryptography, infrastructure or pipelines should receive closer scrutiny. Autonomy could follow the same principle, with platforms restricting an agent's freedom for high-risk tasks while allowing more independence for routine work. Future studies are needed to determine whether these measures preserve productivity and whether developers find ways around them. Longer-term studies should also examine how AI-assisted work affects review skills over time and what security knowledge newcomers to vibe coding need. Users who build applications entirely through conversation may be unable to review the generated code themselves. They therefore need secure defaults and automatic checks before an application is published. 

\subsection{Accountability at Scale}\label{sec:ch-accountability}
Accountability raises a different challenge. Standards such as ISO/IEC~27001 and SOC~2 generally assume that people are involved in writing, reviewing and approving changes, while newer regulations impose security obligations on software makers regardless of how their software was developed \cite{cra}. When agents operate at L4 or L5, it becomes unclear who is responsible for vulnerabilities that no person has examined and how products that are largely generated can demonstrate compliance. Provenance records offer a technical starting point, but audit requirements and the allocation of liability still need agreement among standards bodies and regulators~\cite{elgendy2026responsible, ARR}.

\subsection{An Evaluation Science for Vibe Coding Security}\label{sec:ch-evaluation}
Current evaluations do not fully reflect the security risks of vibe coding. As shown in Section~\ref{sec:taxonomy}, most benchmarks assess individual turns against static attackers, focusing on models fixing human-written code rather than generating code themselves. They also overlook multi-step attacks, repeated sessions and shared platforms. Without more realistic evaluations, it will be difficult to determine whether the proposed defences work in practice. 

Future research should therefore develop a shared benchmark that evolves with new models and agents and reports results separately for each autonomy level, ensuring that a defence shown to work at L1 is not assumed to work at L4. The benchmark should cover realistic repository tasks involving CI and cloud services, attacks that spread across multiple files, compound attack chains, accidental failures as well as deliberate attacks, and repeated tests across model versions. Alongside more realistic benchmarks, studies need common metrics to enable meaningful comparisons. Current evaluations report attack success rates, vulnerability counts, compliance rates and deployability scores that are difficult to compare. 

\begin{table}[t]
\centering
\caption{Research agenda for secure vibe coding. Each challenge lists the Section~\ref{sec:taxonomy} gaps it addresses, the main research directions and established foundations to build on. Nature: E = engineering work with known methods; F = fundamental research.}
\label{tab:agenda}
\scriptsize
\setlength{\tabcolsep}{3pt}
\renewcommand{\arraystretch}{1.2}
\begin{tabularx}{\textwidth}{@{}>{\raggedright\arraybackslash}p{2.3cm} >{\raggedright\arraybackslash}p{2.4cm} >{\raggedright\arraybackslash}X >{\raggedright\arraybackslash}p{2.6cm} c@{}}
\toprule
Challenge & Gaps addressed (\S) & Research directions & Foundations to build on & Nature \\
\midrule
Enforcing limits on agent authority (\ref{sec:ch-authority})
& G1, G8, G9, G15, G17
& Task-based permissions; runtime checks on network, approval gaps and tokens; short-lived credentials; handling unsafe developer instructions
& Reference monitors; capability-based security; scoped tokens
& F/E \\
End-to-end provenance and information flow (\ref{sec:ch-provenance})
& G5, G18, G20 
& Labels that survive model rewriting; per-change provenance records; source attribution on shared platforms
& Information-flow control; SBOM, SLSA and in-toto
& F/E \\
Integrity of persistent agent state (\ref{sec:ch-state})
& G2, G4, G21
& Versioned memory with expiry and rollback; regular re-checking; proven removal; per-tenant separation
& Database versioning; tamper-evident logs
& E/F \\
Assurance of the model supply chain (\ref{sec:ch-model})
& G3 
& Measured resistance to poisoning; signed weights and training data; checkable backdoor removal
& Reproducible and signed builds; certified robustness
& F \\
Security by construction (\ref{sec:ch-construction})
& G6, G7, G13, G14 
& Asking for or inferring security requirements; API grounding; changes submitted with checkable evidence; verified self-healing
& Proof-carrying code; design by contract; policy as code
& E/F \\
Vulnerability lifecycle of generated code (\ref{sec:ch-lifecycle})
& New; builds on G6, G20
& Model-level advisories; tracing affected code; coordinated repair
& CVE and CWE processes; clone detection
& E/F \\
A trustworthy agent ecosystem (\ref{sec:ch-ecosystem})
& G10, G11, G12, G19 
& Pre-install checks; signed, versioned tool descriptions; agent identity; delegation limits; diverse reviewers
& TUF and Sigstore; app permission models
& E \\
Securing the vibe coding toolchain (\ref{sec:ch-toolchain})
& New; builds on G9, G10, G16
& Signed instruction files; permission prompt design; security testing across model versions
& Browser-extension security; regression testing
& E \\
Human oversight and accountability (\ref{sec:ch-human})
& G16 
& Risk-based review effort; adjustable autonomy; field and long-term studies; secure defaults for non-programmers; audit and liability rules
& Alarm-fatigue research; separation of duties
& E/F \\
Accountability at scale (\ref{sec:ch-accountability}) & New; builds on G16, G18, G20 & Audit evidence from provenance records; allocating liability; compliance for agent-made changes & ISO/IEC 27001, SOC 2, EU CRA; SLSA attestations & E/F\\
Evaluation science (\ref{sec:ch-evaluation})
& Cross-cutting (all of \S\ref{sec:taxonomy})
& Shared benchmark by autonomy level; adaptive attackers and compound chains; common metrics including cost; incident database
& CVE programme; OSS-Fuzz
& E \\
\bottomrule
\end{tabularx}
\vspace{-1em}
\end{table}

\section{Conclusion}\label{sec:conclusion}
In this survey, we examined vibe coding security through the full workflow, from the developer's prompt and the model's decisions to tool use, deployment, and production. This view shows that the central risk lies not in any single attack, but in the growing authority of AI agents and the limited oversight of their actions. Familiar weaknesses, including prompt injection, poisoned sources, fake packages, and secret leakage, can lead to direct harm when agents act on them and can combine into attack chains that individual defences fail to stop. Our review shows that most existing defences and benchmarks still assume assistants that only make suggestions. By linking threats to their unresolved gaps and bringing these gaps together into a research agenda, we hope this survey provides a shared basis for researchers, tool builders, and practitioners for making vibe coding safe to use at the level of autonomy it is quickly reaching.

\myparagraph{Threats to Validity} Several factors limit the conclusions of this survey. First, the field changes faster than peer review, so our evidence includes preprints, vendor reports and disclosed incidents, and some systems may have changed since our search ended. Second, our search terms and selection criteria may have missed relevant work that uses different terms for vibe coding or AI coding agents, although forward and backward citation tracking reduces this risk. Third, classifying studies into threat classes and assessing how each risk changes with autonomy involved our own judgement, other researchers might group some threats differently. Finally, we focus on security and do not cover related concerns such as code quality, maintainability or licensing, even though these can share similar causes.

% % ============================================================
% ACKNOWLEDGMENTS
% % ============================================================
\begin{acks}
This research is supported by the CyberASAP project id:10193042, VibeCheck: Intent-Aware Security Assurance for Vibe Coding.
\end{acks}

% ============================================================
% REFERENCES
% ============================================================
\bibliographystyle{ACM-Reference-Format}

\bibliography{main}

% ============================================================
% % APPENDIX (OPTIONAL)
% % ============================================================
\appendix
\section{Appendix A: Boolean Search Query}\label{sec:AppendixA}

The following Boolean search query was used as the basis for searching the ACM Digital Library, IEEE Xplore, USENIX, Scopus, Google Scholar, and arXiv. Minor syntactic modifications were applied where required to accommodate the query language supported by individual digital libraries while preserving the underlying search semantics.

% \begin{lstlisting}[language={},caption={Boolean search query used for literature retrieval},label={lst:searchquery}]
% (
%   "vibe coding" OR
%   "natural language programming" OR
%   "prompt-based programming" OR
%   "conversational programming"
% )
% AND
% (
%   "AI-assisted code generation" OR
%   "AI code generation" OR
%   "large language model" OR
%   "program synthesis" OR
%   "GitHub Copilot"
% )
% AND
% (
%   security OR
%   vulnerability OR
%   vulnerabilities OR
%   "software security" OR
%   "prompt injection"
% )
% \end{lstlisting}

\begin{tcolorbox}[
title={Boolean Search Query},
colback=gray!5,
colframe=black,
boxrule=0.6pt,
arc=2mm,
left=2mm,
right=2mm,
top=1mm,
bottom=1mm]

\ttfamily
\scriptsize
\begin{verbatim}
( "vibe coding" OR "AI coding assistant" OR "coding agent" OR
  "code generation" OR "GitHub Copilot" OR "LLM-generated code" )
AND
( "large language model" OR "LLM" OR "agent" OR "retrieval-augmented" OR
  "Model Context Protocol" OR "CI/CD" OR "infrastructure as code" )
AND
( "security" OR "vulnerability" OR "attack" OR "prompt injection" OR
  "poisoning" OR "supply chain" OR "privacy" )
\end{verbatim}
% (
%   ``vibe coding'' OR

%   ``natural language programming'' OR

%   ``prompt-based programming'' OR

%   ``conversational programming''
% )

% AND

% (
%   ``AI-assisted code generation'' OR

%   ``AI code generation'' OR

%   ``large language model'' OR

%   ``program synthesis'' OR

%   ``GitHub Copilot''
% )

% AND

% (
%   ``security'' OR

%   ``vulnerability'' OR

%   ``vulnerabilities'' OR

%   ``software security'' OR

%   ``prompt injection''
% )

\end{tcolorbox}

% \section{Additional Materials}
% % Extra tables, proofs, etc.

\end{document}